\documentclass[twocolumn,twoside]{IEEEtran}

\IEEEoverridecommandlockouts

\usepackage{booktabs} 
\usepackage{rotating}

\usepackage{graphicx}
\usepackage{subfigure}
\usepackage{amsmath}
\usepackage{amssymb}
\usepackage{array}
\usepackage{multirow}
\usepackage{color,colortbl}
\usepackage{hyperref}
\usepackage{cite}
\let\labelindent\relax
\usepackage{enumitem}
\usepackage{pifont}
\newcommand{\xmark}{\ding{55}}%
\newcolumntype{M}[1]{>{\centering\arraybackslash}m{#1}}

\begin{document}

\bibliographystyle{IEEEtran}
\title{Big Data Analytics for Large Scale Wireless Networks: Challenges and Opportunities}

\author{
Hong-Ning Dai,\thanks{H.-N. Dai is with Faculty of Information Technology, Macau University of Science and Technology, Macau. E-mail: hndai@ieee.org.}
Raymond Chi-Wing Wong,\thanks{R. C.-W. Wong is with Department of Computer Science and Engineering (CSE)
the Hong Kong University of Science and Technology (HKUST), Hong Kong. Email: raywong@cse.ust.hk.}
Hao Wang,\thanks{H. Wang is with Faculty of Engineering and Natural Sciences, Norwegian University of Science \& Technology in Aalesund, Norway. Email: hawa@ntnu.no}
Zibin Zheng, \thanks{Z. Zheng is with School of Data and Computer Science, Sun Yat-sen University, Guangzhou, China. Email: zibin.gil@gmail.com}
Athanasios V. Vasilakos \thanks{A. V. Vasilakos is with Department of Computer Science, Electrical and Space Engineering, Lulea University of Technology, 97187, Lulea, Sweden. Email: vasilako@ath.forthnet.gr.}
}

\maketitle


\begin{abstract}

The wide proliferation of various wireless communication systems and wireless devices has led to the arrival of big data era in large scale wireless networks. Big data of large scale wireless networks has the key features of wide variety, high volume, real-time velocity and huge value leading to the unique research challenges that are different from existing computing systems. In this paper, we present a survey of the state-of-art big data analytics (BDA) approaches for large scale wireless networks. In particular, we categorize the life cycle of BDA into four consecutive stages: Data Acquisition, Data Preprocessing, Data Storage and Data Analytics. We then present a detailed survey of the technical solutions to the challenges in BDA for large scale wireless networks according to each stage in the life cycle of BDA. Moreover, we discuss the open research issues and outline the future directions in this promising area.

\end{abstract}



  

\section{Introduction}
\label{sub:intro}

In recent years, we have seen the proliferation of wireless communication technologies, which are widely used today across the globe to fulfill the communication needs from the extremely large number of end users. The interconnection of various wireless communication systems together forms a \emph{large scale} wireless networks, where ``large scale'' means the high density of network stations (or nodes) and the large coverage area. Meanwhile, there is a surge of big volume of mobile data traffic generated from wireless networks that consist of a wide diversity of wireless devices, such as smart-phones, mobile tablets, laptops, RFID tags, sensors, smart meters and smart appliances. It is predicted that in a report of \cite{cisco:whitepaper17} there is a growth of the mobile data traffic from 10 EB/month (1EB $ = 1\times 10^{18}$ bytes) in 2017 to 49 EB/month in 2021, representing that we are entering a ``\emph{big data era}'' \cite{LCui:IEEENet2016}. 

Essentially, big data has the following salient features called ``4Vs'' differentiating it from other concepts, such as ``very large data'', ``large volume data'' and ``massive data'' \cite{IBM:2011}:
\begin{enumerate}
\item \emph{Volume.} The quantity of generated and stored data (usually refer to the data volume from Terabytes to Petabytes);
\item \emph{Variety.} The type and nature of the data (structured, semi-structured, unstructured, text and multimedia);
\item \emph{Velocity.} The speed at which the data is generated and processed to meet the demands (e.g., real time).
\item \emph{Value.} The analytical results based on big data can bring huge both business value and social value.
\end{enumerate}

\begin{table}[t]
\caption{Comparison of this article with existing surveys}
\footnotesize
\renewcommand{\arraystretch}{2.15}
\begin{tabular}{|c|c|c|}
\hline
\textbf{Research issues} & \textbf{References} & \textbf{This survey}\\
\hline
General BDA & \cite{Wu:TKDE14} \cite{Hu:IEEEAccess14} \cite{chen2014big} & \checkmark \\
\hline
Wireless Sensor Networks (WSN) & \cite{Alsheikh:2014} & \checkmark \\
\hline
Mobile Communication Networks & \cite{SBi:IEEEComMag15} \cite{CJiang:IEEEWC2017} \cite{kibria:2017bigdata} \cite{Qian:JCIN2017} & \checkmark \\
\hline
Vehicular networks & \xmark & \checkmark \\
\hline
Mobile Social Networks & \xmark & \checkmark\\
\hline
Internet of Things (IoT) & \cite{Qian:JCIN2017}  & \checkmark \\
\hline	
\end{tabular}
\label{tab:comp}
\end{table}

Although there are other two 'V's, i.e., ``Variability'' and ``Veracity'' \cite{Hilbert:bigdata16}, we mainly use the above ``4Vs'' to describe big data generated from wireless networks. Since there are various types of large scale wireless networks, we only enumerate several exemplary networks including mobile communication networks, vehicular networks, mobile social networks, and Internet of Things (IoT). The wireless devices include not only various wired interfaces and wireless interfaces but also sensors consisting of temperature sensor, light sensor, acoustic sensor, vibration sensor, chemical sensor, accelerator and RFID tags \cite{WangLiu:CST11}, which can generate high volume data in real-time fashion. In summary, big data generated from large scale wireless networks is often featured with wide variety, high volume, real-time velocity and huge value.

The growth of big data in large scale wireless networks brings not only the \emph{challenges} in designing scalable wireless networks but also the \emph{value}, which is beneficial to many areas, such as network operation, network management, network security, network optimization, intelligent traffic system, logistic management and social behavior study. It requires \emph{big data analytics} (BDA) dedicated for large scale wireless networks to harness the benefits. Data generated from large scale wireless networks should be collected, filtered, stored and analyzed until the ``value'' is extracted.

\begin{figure*}[t]
\centering
\includegraphics[width=15cm]{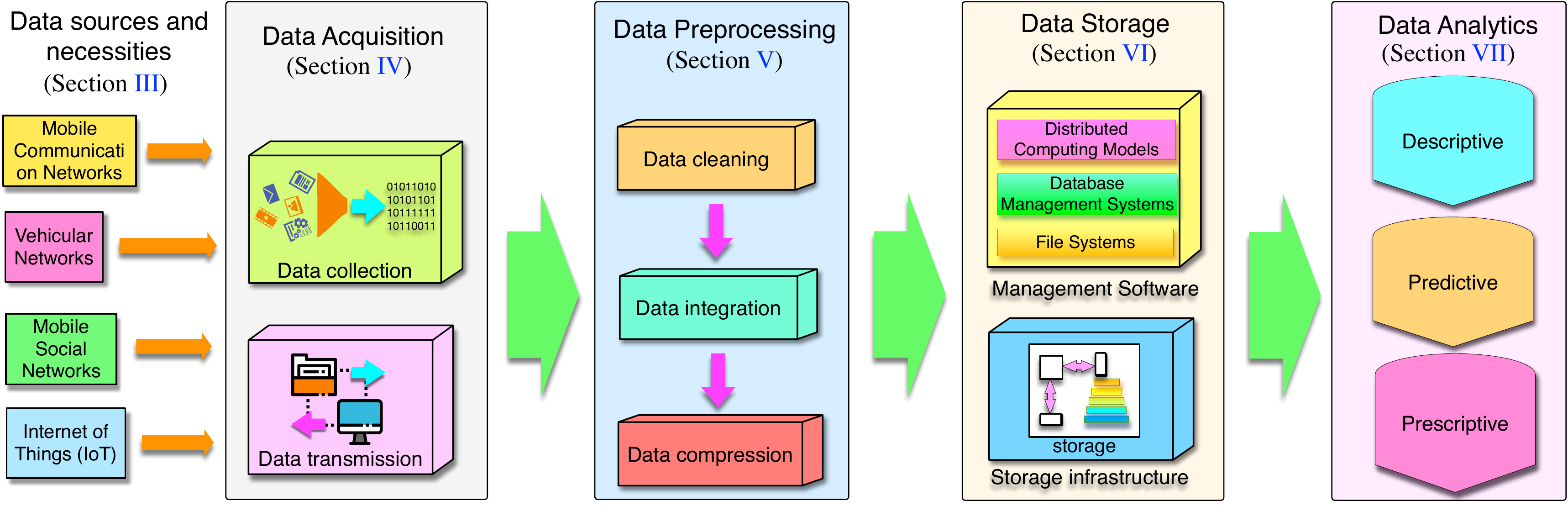}
\caption{Life cycle of Big Data Analytics in large scale wireless networks}
\label{fig:BDA-lifecycle}
\end{figure*}

\subsection{Comparisons between this paper and existing surveys}

There are a number of studies related to BDA including data mining, machine learning and distributed computing. For example, a survey on big data processing models from the \emph{data mining perspective} is presented in \cite{Wu:TKDE14}. The study of \cite{Hu:IEEEAccess14} presents a tutorial on BDA from the \emph{scalability of BDA platforms}. Ref. \cite{chen2014big} gives a survey on BDA from the aspect of \emph{enabling technologies}. However, these previous surveys concentrated on BDA for general computing systems (e.g., data warehouse) and are not dedicated for large scale wireless networks, which have different features. For example, data generated in wireless networks is usually in real-time fashion and in heterogeneous types. Therefore, the conventional BDA approaches in general computing systems cannot be applicable to large scale wireless networks.

Recently, several surveys on BDA in wireless networks have been published. The paper \cite{Alsheikh:2014} presents a survey on using machine learning methods in wireless sensor networks (WSNs). The study of \cite{SBi:IEEEComMag15} gives a short overview on big data analytics in wireless communication systems. In \cite{CJiang:IEEEWC2017}, an overview on machine learning in next-generation wireless networks is presented. Ref. \cite{kibria:2017bigdata} presents an overview on BDA and artificial intelligence in next-generation wireless networks. Qian et al. \cite{Qian:JCIN2017} survey a limited number of studies from data, transmission, network and application layers in wireless networks including communication networks and Internet of Things (IoT). 

However, these studies are \emph{too specific} to a certain type of wireless networks (either WSNs or mobile communication networks). Essentially, different wireless networks being featured of heterogeneous data types require different BDA approaches. For instance, vehicular networks have less computational and energy constraints than wireless sensor networks. In contrast to the above-mentioned surveys, we attempt to provide an in-depth survey on BDA for large scale wireless networks with the inclusion of up-to-date studies. Our survey has a good horizontal and vertical coverage of research issues in BDA for large scale wireless networks. In the horizontal dimension, we mainly focus on four phases in the life cycle of BDA. In the vertical dimension, we consider four representative wireless networks (namely WSNs, Mobile Communication Networks, Vehicular networks and IoT). Table \ref{tab:comp} highlights the differences between this survey and other existing surveys.

\subsection{Contributions}

We first conduct a comprehensive literature collection and analytics with consideration of timeliness, relevance and quality. The research methodology is presented in Section \ref{sec:methodology}. We then introduce typical data sources of large scale wireless networks and discuss the necessities BDA for large scale wireless networks in Section \ref{sec:bigdata}. 

The core contribution of this paper is to present the state-of-the-art of BDA in the context of large scale wireless networks in two dimensions: 1) life cycle of BDA and 2) different types of wireless networks. In order to give readers a clear roadmap about the BDA procedures, we introduce the life cycle of BDA. As shown in Fig. \ref{fig:BDA-lifecycle}, we categorize the life cycle of BDA into four consecutive stages: Data Acquisition, Data Preprocessing, Data Storage and Data Analytics. Note that the data flow along the above four stages may not strictly go forward. In other words, there might be some backward links from one stage to the preceding stage. For example, the data flow in the data analytics stage may go back to the data storage stage since some statistic modeling algorithms require the comparison of the current data with the historical data. It is also worth mentioning that there are other taxonomies of the phases of BDA proposed for other computing systems \cite{Hu:IEEEAccess14,Casado:CCPE2015}. In this paper, we categorize the life cycle of BDA into the above four stages since this categorization can accurately capture the key features of BDA in large scale wireless networks, which are significantly different from other computing systems. We next briefly describe them as follows.

\begin{figure*}[t]
\centering
\includegraphics[width=15cm]{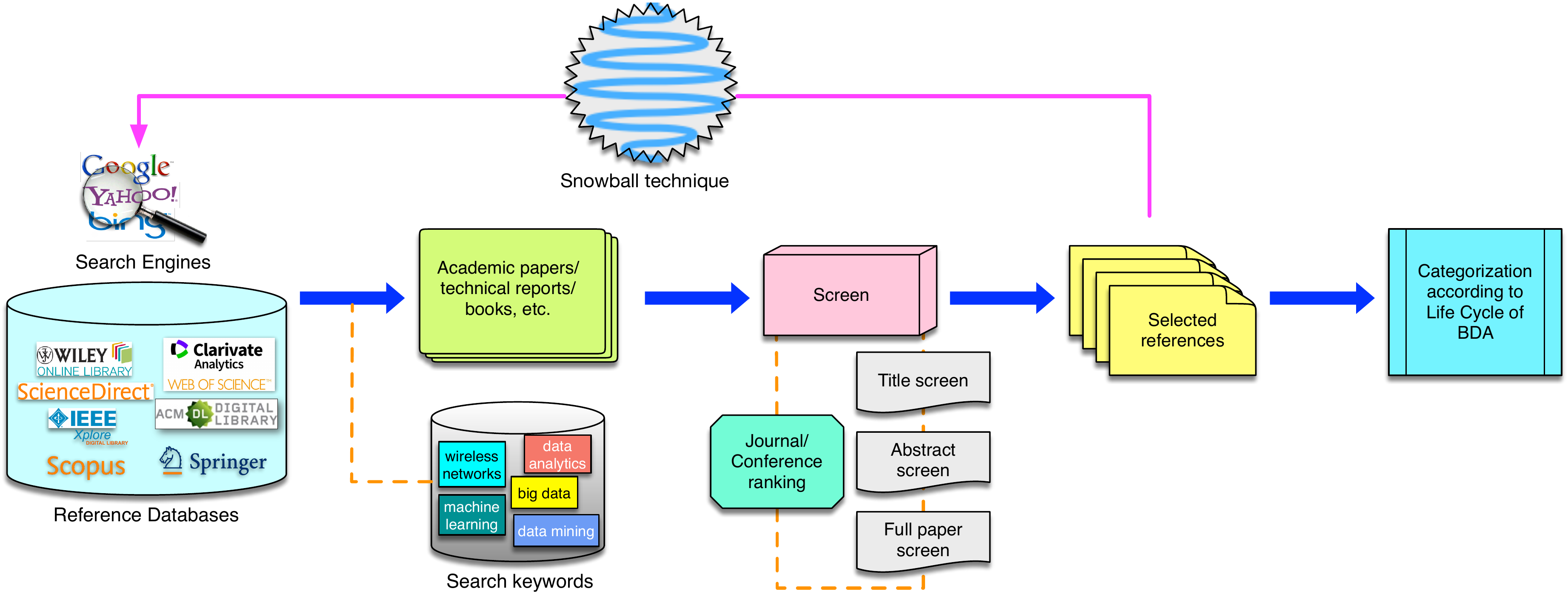}
\caption{Methodology adopted in this article}
\label{fig:methodology}
\end{figure*}

\begin{enumerate}
\item[1.] \emph{Data acquisition.} Data acquisition consists of data collection and data transmission. In particular, data collection involves acquiring raw data from various data sources with dedicated data collection technologies, for example, reading RFID tags by RFID readers in IoT. Then, the data is then transmitted to the data storage system via wired or wireless networks. Details about data acquisition are given in Section \ref{sec:acquisition}.

\item[2.] \emph{Data preprocessing.} After collecting raw data, the raw data needs to be preprocessed before keeping them in data storage systems because of the big volume, duplication, uncertainty features of the raw data \cite{LChen:TKDE12}. The typical data preprocessing techniques include data cleaning, data integration and data compression. We present more details on data preprocessing in Section \ref{sec:preprocessing}.

\item[3.] \emph{Data storage.} Data storage refers to the process of storing and managing massive data sets. We divide the data storage system into two layers: storage infrastructure and data management software. The infrastructure not only includes the storage devices but also the network devices connecting the storage devices together. In addition to the networked storage devices, data management software is also necessary to the data storage system. Details about data storage are given in Section \ref{sec:storage}.

\item[4.] \emph{Data analytics.} In this phase, various data analytics schemes are used to extract valuable information from the massive data sets. We roughly categorize the data analytics schemes into three types: (i) descriptive analytics, (ii) predictive analytics and (iii) prescriptive analytics. Details on data analytics are presented in Section \ref{sec:analysis}.
\end{enumerate}

It is worth mentioning that we also consider different types of wireless networks in each of the above stages. In addition, we present some open research issues and discuss the future directions in this promising area in Section \ref{sec:open}. Finally, we conclude this paper in Section \ref{sec:conc}.

\section{Research methodology}
\label{sec:methodology}

\subsection{Reference databases and search criteria}

Fig. \ref{fig:methodology} gives the schematic illustration of the methodology adopted in this paper. In particular, we query seven reference databases to obtain the relevant articles/books: 1) ACM Digital Library, 2) Claviate Analytics Web of Science, 3) IEEE Xplore, 4) ScienceDirect, 5) Scopus, 6) SpringerLink and 7) Wiley Online Library. Moreover, we also exploit mainstream search engines to obtain the relevant literature.

Furthermore, in order to include relevant papers as many as possible, we establish a keyword-dataset consisting of keywords and their synonyms. For example, Internet of things may be relevant to wireless sensor networks, Machine-to-Machine communications, cyber-physical systems, smart city, RFID, etc. We search relevant literature according to the search string with the ``OR'' connection of keywords and their synonyms. Table \ref{tab:keywords} gives the representative search keywords and their synonyms.

\subsection{Data extraction}

In the second step, we then read titles and abstracts for initial screening. If necessary, we will read other parts of some papers. It is worth mentioning that we are selective when including relevant high-quality papers while we exclude irrelevant and non-peer-reviewed papers (e.g., papers published in predatory journals). Therefore, we apply journal/conference rankings in the screening process (e.g., Scientific Journal Rankings, Journal Citation Reports, Excellence in Research for Australia, etc.). Moreover, to reflect timeliness of this research area, we choose time window from 2002 to 2018 with an exception of papers related to background knowledge of specific technologies (e.g., storage technologies as shown in Section \ref{sec:storage}).

\begin{table}[h]
\caption{Representative search keywords and synonyms}
\footnotesize
\renewcommand{\arraystretch}{1.75}
\begin{tabular}{|m{3cm}|m{5cm}|}
\hline
\textbf{Keywords} & \textbf{Synonyms}\\
\hline\hline
Big data analytics & Big data, Massive data, Machine learning, Deep learning, Data mining, Data analysis, Data science, data cleaning, etc.\\
\hline
Mobile communication networks & Wireless networks, Wireless communications, Cellular networks, WiFi, 802.11, Mobile networks, Mobile communications, 5G, 4G, etc. \\
\hline
Mobile social networks & Social networks, Community, Online Social Networks (OSN), Graph mining, Social media, etc. \\
\hline
Vehicular networks & Vehicular technology, Vehicle, Vehicle-to-vehicle (V2V), transportation systems, intelligent transportation systems (ITS), traffic flow, etc. \\
\hline	
Internet of Things & Internet of Things, IoT, Machine-to-Machine (M2M), Wireless Sensor Networks, WSNs, RFID, Cyber Physical Systems, etc. \\
\hline
\end{tabular}
\label{tab:keywords}
\end{table}

\begin{table*}[t]
\caption{Data elements extracted from the references}
\footnotesize
\centering
\renewcommand{\arraystretch}{2.15}
\begin{tabular}{|M{1.5cm}|m{4.5cm}|m{10cm}|}
\hline
\textbf{No.} & \textbf{Element} & \textbf{Description}\\
\hline\hline
1 & Bibliographic information & Authors, title, publication year, source of the paper\\
\hline
2 &Type of paper & journal, conference, book, technical report, white paper\\
\hline
3 &Categorization & four stages in BDA life cycle and four types of wireless networks\\
\hline
4 &Novelty & Are new approaches proposed? \\
\hline
5 &Validation & Have experimental results validated the observations?\\
\hline 
6 &Research challenges & Have research challenges been addressed?\\
\hline
7 &Open directions & Are any implications given? Are any open research issues raised?\\
\hline	
\end{tabular}
\label{tab:extraction}
\end{table*}

In the third step, we then thoroughly review the articles obtained from initial screening and identify the relevance to big data analytics in wireless networks. In addition, we also extend the systematic literature review by using snowballing technique (as shown in Fig. \ref{fig:methodology}). The main idea of snowballing technique is to use the references or citations of a paper to further include other relevant studies. The advantages of snowballing technique include: i) complimenting traditional systematic review methods, ii) locating hidden while important literature and iii) focusing specific relevant topics \cite{wohlin2014guidelines}. The usage of references is named as the backward snowballing method (BSM) while the usage of citations is named as the forward snowballing method (FSM). In this article, we use both BSM and FSM. Finally, we obtain 249 references after this step. Table \ref{tab:extraction} gives the reference extraction guideline, which include major data elements when we select the articles.

\subsection{Distribution of references}

\begin{figure}[h]
	\centering
	\subfigure[\scriptsize Publication years (horizontal axis) vs No. of selected papers (vertical axis).]{
		\includegraphics[width=8cm]{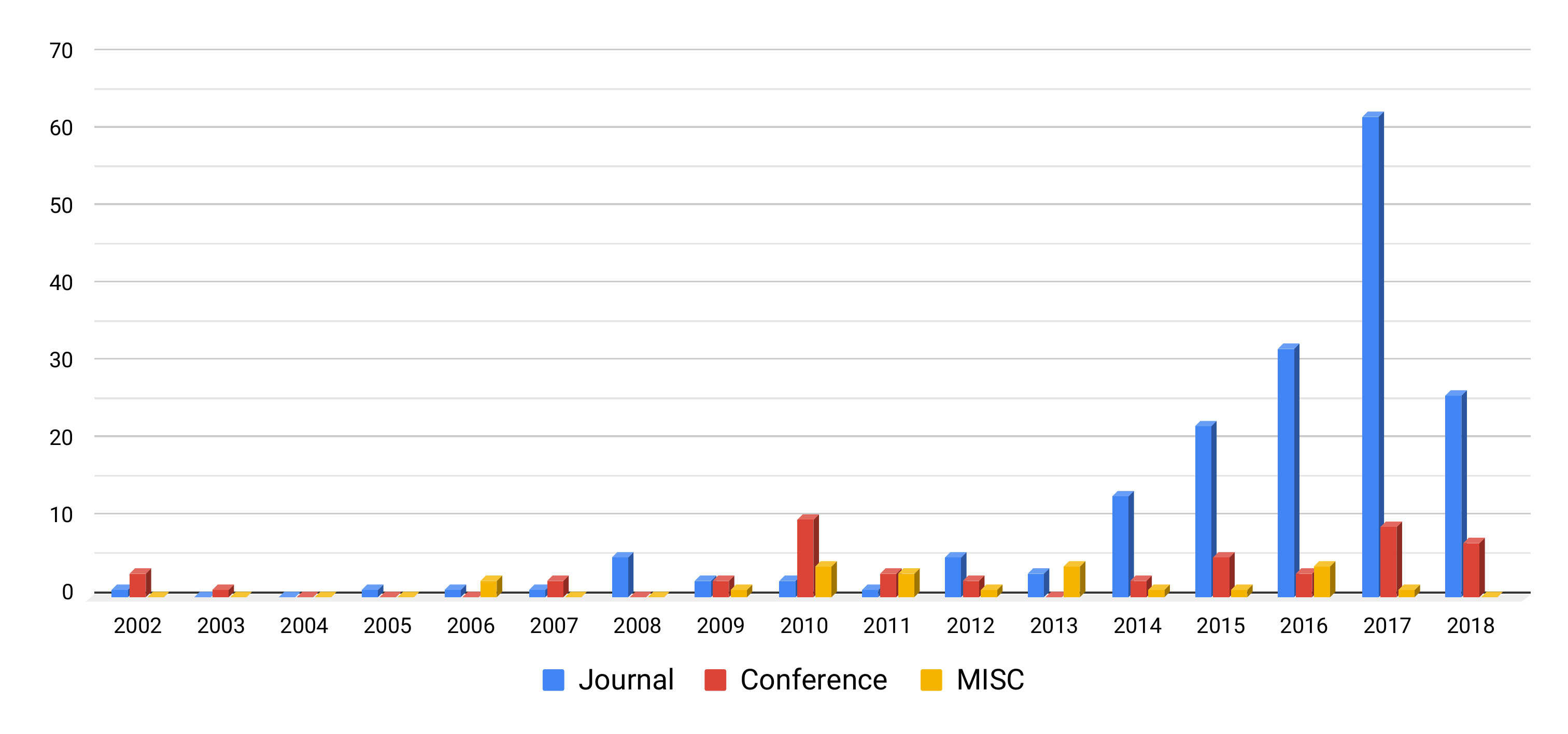}\label{Fig:by-year}}
	\subfigure[\scriptsize Publication types]{
		\includegraphics[width=6cm]{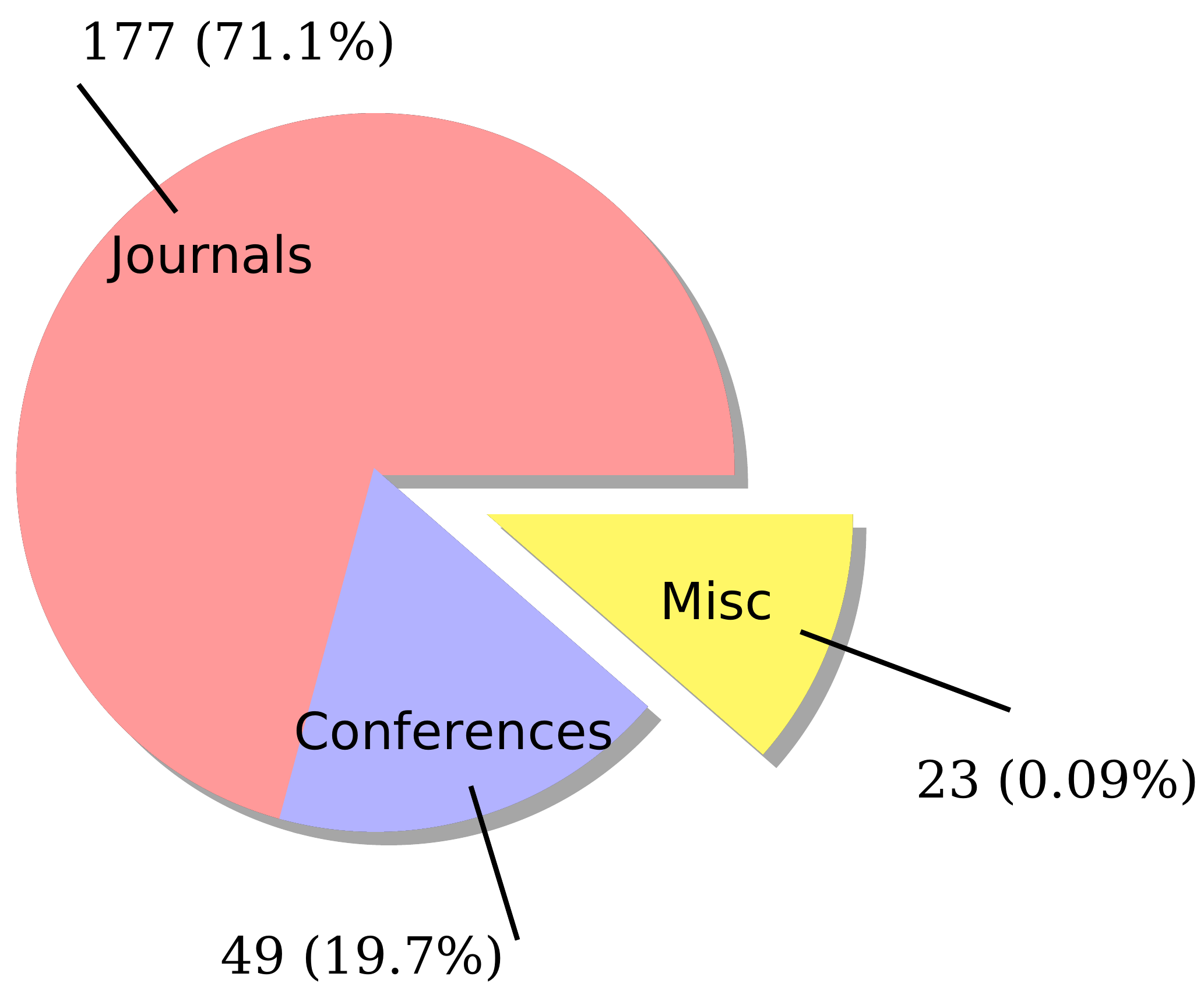}\label{Fig:by-type}}
	\caption{Distribution of selected references}
	\label{Fig:dist_ref}
\end{figure}

\begin{figure*}[t]
	\centering
	\subfigure[\scriptsize Conference articles vs publishers]{
		\includegraphics[width=8.0cm]{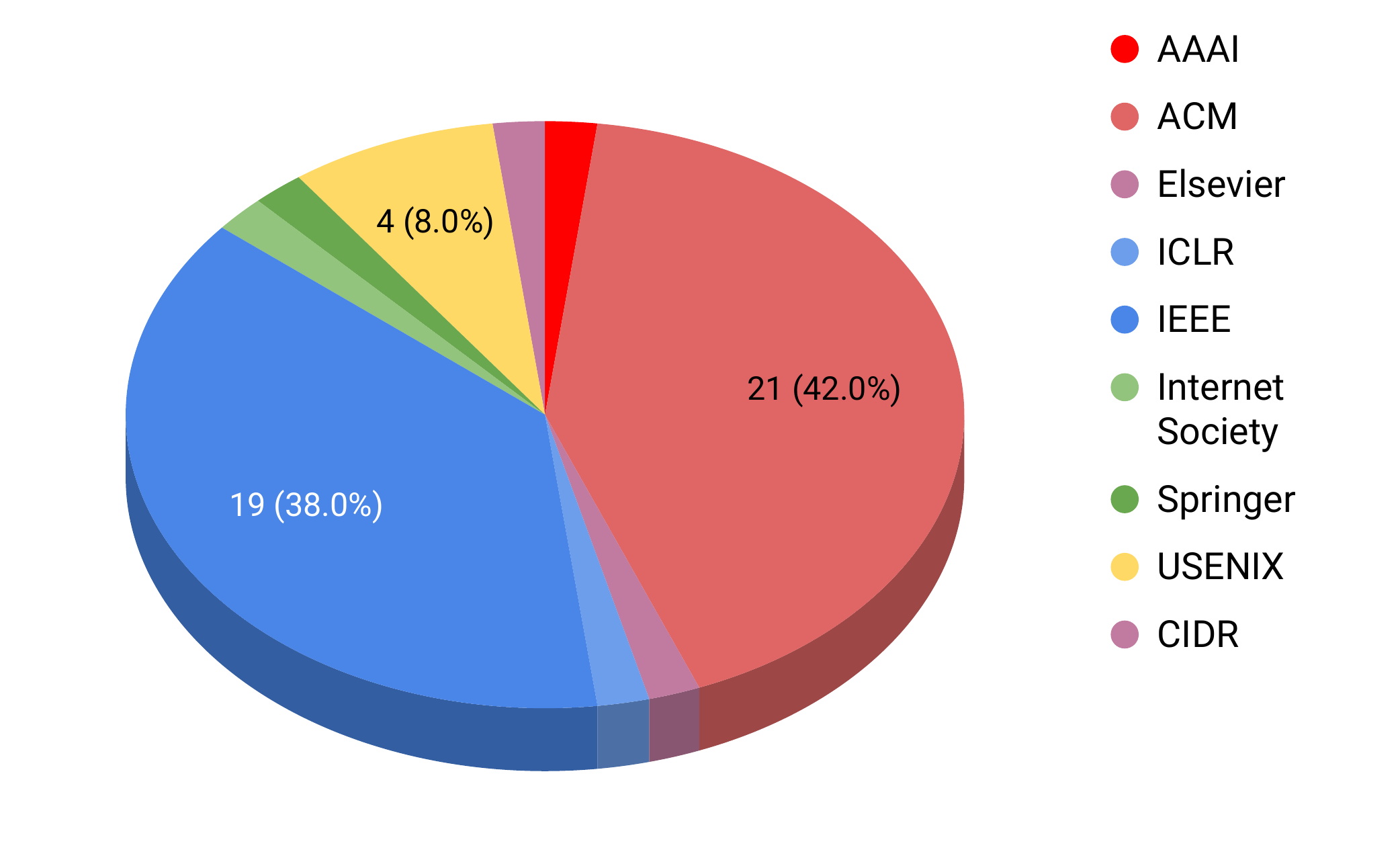}\label{Fig:by-conference}}
	\subfigure[\scriptsize Journal articles vs publishers]{
		\includegraphics[width=8.0cm]{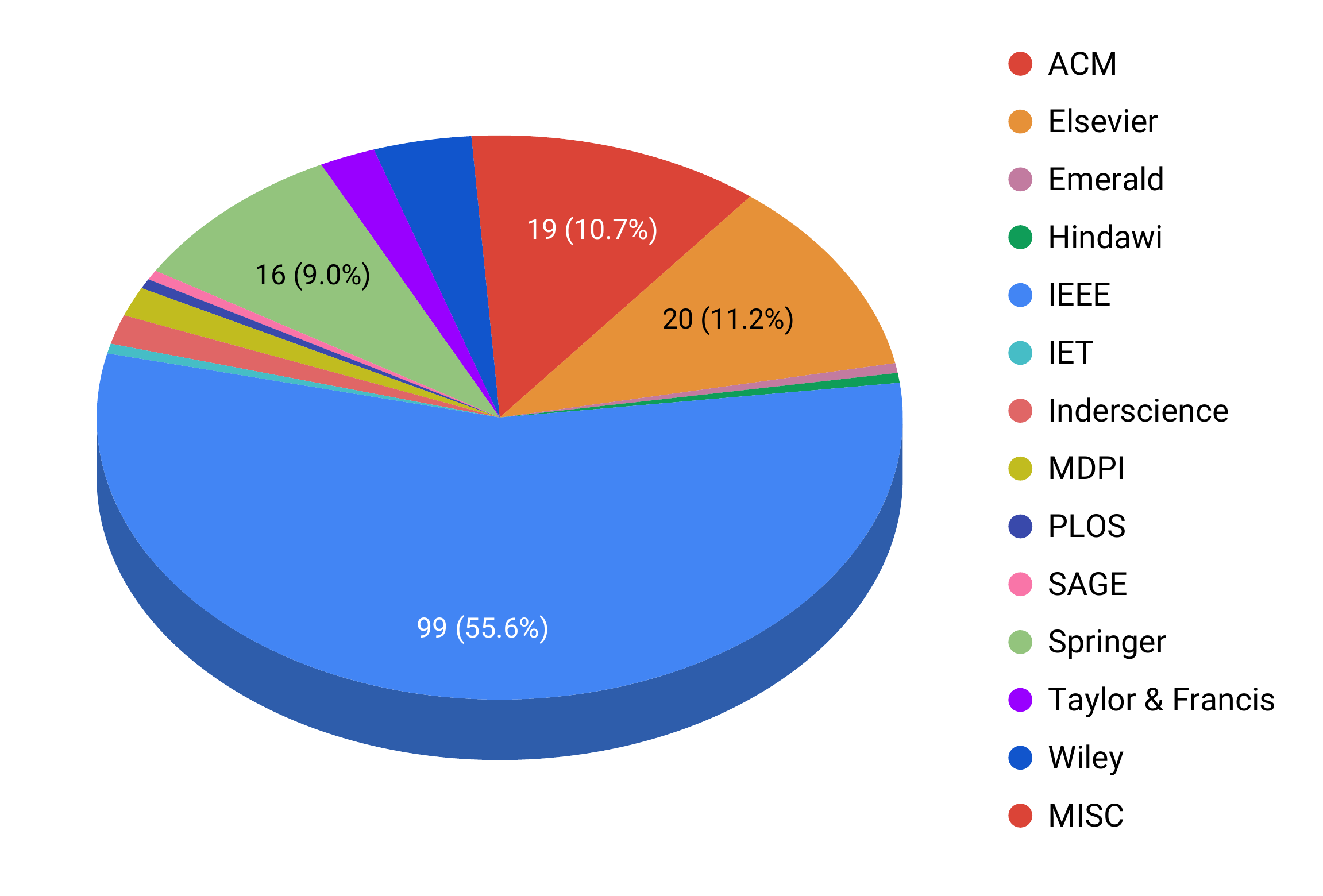}\label{Fig:by-journal}}
	\caption{Proportions of conference and journal articles vs publishers}
	\label{Fig:prop_publisher}
\end{figure*}

Fig. \ref{Fig:dist_ref} presents an overview of the selected papers in terms of their publication years and types. In particular, Fig. \ref{Fig:by-year} shows the distribution of the 249 papers from 2002-2018. We observe that there are few papers between 2002 and 2009 while the number of papers has grown steadily from 2013 to 2018. In 2017, there are 72 papers published, which nearly count for 31\% of all the selected papers. It implies that \emph{big data analytics in wireless networks is becoming a hot topic}. According to the published venues, we further categorize papers into the following types: 1) journal, 2) conference/symposium/workshop, 3) miscellaneous (including books, technical reports, etc.). Fig. \ref{Fig:by-type} shows the distribution of papers according to the types of papers. We notice that there are more journal papers than other types of papers. It may owe to the fact that we prefer journal articles to other types of papers during the screening process since journal papers typically contain more technical details than other types of papers though we admit that many top-tier conference (such as USENIX NSDI) papers also contain adequate technical details.

Since journal and conference articles occupy the largest proportion (i.e., nearly 90\%) among all the references, we further analyze the proportions of conference/journal articles versus publishers. It is shown in Fig. \ref{Fig:by-conference} that ACM has published the majority of conference articles (42\%) followed by IEEE (38\%); both of them occupy nearly 80\% of total number of conference articles. With regard to journal articles, IEEE has the largest proportion of journal articles (55.5\%) among all the publishers followed by Elsevier (11.2\%), ACM (10.7\%) and Springer (9\%) as shown in Fig. \ref{Fig:by-journal}.

\begin{table*}[h]
\caption{Summary of the topics in BDA of wireless networks}
\centering
\footnotesize
\renewcommand{\arraystretch}{2.5}
\begin{tabular}{|M{3.0cm}|M{1.8cm}|M{2.0cm}|M{1.8cm}|M{1.8cm}|M{1.5cm}|M{1.0cm}|}
\hline
& \textbf{Data Acquisition} & \textbf{Data Preprocessing}  & \textbf{Data Storage} & \textbf{Data Analytics} & \textbf{Misc} & \textbf{Sub-total} \\
\hline
\hline
General topics of BDA & 6 & 12 & 48 & 23 & 24& 113 \\ 
\hline

Mobile communication networks & 12 &  4 &  3 & 5 & 11 & 35\\
\hline
Mobile social networks & 6 & 5 & 6 & 8 & 4 & 29\\
\hline
Vehicular networks &6 & 4 & 6 & 4 & 7 & 27\\
\hline
Internet of Things & 12 & 11 & 10 &  7& 5 & 45\\

\hline
\multicolumn{6}{|c|}{\it Total}  & 249 \\
\hline
\end{tabular}
\label{tab:popular-topic}
\end{table*}

Table \ref{tab:popular-topic} summarizes all the topics in BDA for wireless networks. We categorize the topics mainly according to  four types of wireless networks: 1) mobile communication networks, 2) mobile social networks, 3) vehicular networks and 4) Internet of Things. All of them counts for the largest proportion (i.e., 136 papers) among all the references. We observe from Table \ref{tab:popular-topic} that ``Internet of Things'' is one of the hottest topics and it has received extensive attention recently. In addition to four types of wireless networks, we put the remaining 113 papers in the category of general topics of BDA; this class includes references that cover the general topics of BDA ranging from data sources, data acquisition, data preprocessing, data storage and data analytics. 

In addition to the horizontal categorization, Table \ref{tab:popular-topic} also shows the vertical categorization of the topics according to the four stages in the life cycle of BDA, i.e., data acquisition, data preprocessing, data storage and data analytics. It is worth mentioning that we put other relevant topics  such as data sources, necessities and challenges of BDA into \emph{Misc} (i.e., miscellaneous) type. We try to include as many synonyms of key terms as possible when we analyze the papers. Moreover, we also avoid double counting when the content of a paper covers two different types of topics (in this case, we choose the closest topic for this paper after thoroughly reviewing it).   



\section{Data sources and necessities of BDA}
\label{sec:bigdata}

 In this section, we first introduce several typical examples of big data sources of large scale wireless networks in Section \ref{subsec:datasource}. These data sources include: (1) Mobile Communication Networks as introduced in Section \ref{subsec:mobilenet}, (2) Vehicular networks as introduced in Section \ref{subsec:vanet}, (3) Mobile Social Networks as introduced in Section \ref{subsec:msn} and (4) Internet of Things (IoT) as introduced in Section \ref{subsec:iot}. We next discuss necessities of BDA in wireless networks in Section \ref{subsec:necessity}.

\subsection{Data sources}
\label{subsec:datasource}

\subsubsection{Mobile Communication Networks}
\label{subsec:mobilenet}


Mobile communication networks are experiencing a shift from single, simple, low data-rate transmission to multiple, complex, high data-rate transmission with the evolution of mobile communication systems. For example, the downlink data rate of a wireless device in the 5G mobile networks is greater than 20 Gbps, which is about 100 times of that in 4G mobile systems \cite{Shafi:JSAC2017}. This shift also exhibits in the wide diversity of data types (e.g., 4k video streams, high fidelity audio, RAW pictures, heart rate, spatial and temporal data in 5G mobile networks). Note that the data generated by cellular networks are not only user data but also system-level data (including cell-level, core-network-level data, etc.) \cite{WXu:WirelessComMag2018}. 

With the growing demands of the bandwidth-ravenous applications, several new wireless access architectures, such as coordinated multipoint (CoMP) \cite{Irmer:IEEECST17}, massive MIMO \cite{Molisch:ComMag2017}, Non-Orthogonal Multiple Access (NOMA) \cite{Shin:ComMag2017} and cloud-based radio access network (C-RAN) \cite{Checko:IEEECST15} have been proposed. Besides, there is a trend of the fusion of cellular networks with other wireless networks, such as wireless LAN (WLANs), wireless personal area networks (WPANs), and small-cell networks together to form a heterogeneous network (HetNet) \cite{Mehmeti:TMC2017}.

The proliferation of various coexisting wireless networks in HetNets also results in the wide diversity of data sources. In particular, data sources in HetNets can be categorized into the following types:
\begin{itemize}
\item \emph{Subscriber-related data} contains control data and contextual data. Examples include call setup time, call success rate, call drop rate, signaling, packet jitter, delay, etc. In addition, it also includes subscriber specific data \cite{WXu:WirelessComMag2018}.
\item \emph{Network-related data source} contains both radio (Physical) measurement data and Base Stations (BSs) layer-2 (Link) measurement data, where BSs are referred to macro BSs, micro BSs, pico BSs, femto BSs, WiFi APs, etc. Moreover, it also includes network specific data such as faults, configurations, accounting information and performance information.
\item \emph{Application Data} contains social media, smartphone sensor data, mobility status, locations, weather, etc.
\end{itemize}

\subsubsection{Vehicular Networks}
\label{subsec:vanet}

Vehicular safety and transportation optimization have received extensive attention recently since cars and other private vehicles are playing an important role in our daily life. Vehicular Networks (VNets) were proposed \cite{Cooper:CST2017,MacHardy:CST2018} to fulfill safety and efficiency requirements of Intelligent Transportation Systems (ITS). There are two typical communications in a VNet: vehicle-to-vehicle (V2V) communications and vehicle-to-infrastructure (V2I) communications. Various wireless communication technologies were proposed to support V2V and V2I communications. These technologies include IEEE 802.11 (WLAN or WiFi), IEEE 802.11p Dedicated Short Range Communications (DSRC) and Wireless Access in a Vehicular Environment (WAVE) \cite{Bila:ITS2017} and the aforementioned 2G-4G communication technologies.

VNets provide vehicle drivers and other road users (e.g., road operators and pedestrians) with a wide range of information, which can be used to enhance the road safety, the public security, the traveling comfort of passengers and the efficiency of optimizating traffic flows \cite{Koesdwiady:TVT16}. In particular, we categorize the data sources generated from VNets into the following types.
\begin{itemize}
\item \emph{Traffic flow data} contains vehicular speed, density of vehicles, vehicular flow and traffic bottlenecks, which can be used to design an optimal road network and minimize the traffic congestion \cite{Liu:adhoc2016}. 
\item \emph{Public safety/security data} includes the route information of suspicious vehicles (e.g., conducting a terrorism behavior) and the event messages of emergency vehicles (e.g., an ambulance uses a lane preemptively). 
\item \emph{Vehicular safety warning messages} include intersection collision avoidance, turn signals (left turn or right turn), lane change warning, blind spot warning, etc.
\item \emph{Ride quality monitoring information} includes the roughness of a road surface (affecting the ride quality) and the slipperiness of a road surface (affecting the ride safety) \cite{Liu:TITS2017}.
\item \emph{Location-aware social network information} mainly includes not only the messages or micro-blogs of some emergencies, such as traffic jams, malfunctioning traffic signals and accidents but also the traveling information, such as the location and the prices of the nearest restaurant and petrol stations \cite{Giridhar:2016}. 

\end{itemize}  

The above wide range of data was usually generated from various sensors (such as accelerometer, laser sensors and GPS), cameras, wireless devices of vehicles, smart phones and RFIDs (that are used for re-identification at electronic toll collection transponders). Besides, most of the data is generated in real time and in a time-critical way. For example, a road dangerous warning message could be sent to drivers within several hundred milliseconds \cite{Bila:ITS2017}. 

\begin{table*}[t]
\caption{Summary of data sources of large scale wireless networks}
\centering
\footnotesize
\renewcommand{\arraystretch}{1.5}
\begin{tabular}{|m{2.2cm}|m{1.0cm}|m{4.2cm}|m{1.7cm}|m{3.8cm}|}
\hline
\textbf{Data source} & \textbf{Volume}  &  \textbf{Variety} &  \textbf{Velocity} & \textbf{Value} \\
\hline
\hline
Mobile Communication Networks &  TB & {\it User data}: unstructured, semi-structured   {\it Operator data}: structured  &  very fast (real time) &  User satisfaction, operation efficiency, system reliability \\
\hline
 Vehicular Networks &  TB & structured, semi-structured  &  very fast (real time)  &  Transportation safety, transportation efficiency, ride quality\\
\hline
 Mobile Social Networks &  PB &  structured, semi-structured, unstructured  & fast  &  Personal interests, user behavior, social welfare, demography\\
\hline
 Internet of Things &  TB to PB &  structured, semi-structured, unstructured  &  fast  & Environment protection, industrial productivity, public safety \\
\hline
\end{tabular}
\label{tab:datasources}
\end{table*}

\subsubsection{Mobile Social Networks (MSNs)}
\label{subsec:msn}

Mobile social networks (MSNs) can provide mobile users with various social applications and services due to the proliferation of wireless networks and various mobile devices \cite{QXu:TETC15,ZSu:IEEENet16}. 
There are various data sources generated from MSNs including login information, personal profiles, rating information or interests (e.g., ``likes''), contextual data (e.g., tags, status, location), photos, video, etc. We categorize them into the following types.
\begin{enumerate}
\item \emph{Service Provider-related Data} mainly refers to the data originates from the service usage of social networks. They include the following sub-types: (i) \emph{Login data.} Social network service providers require the prior user authentication to prevent from the identity theft. (ii) \emph{Connection data.} The connections to MSNs result in a large volume of digital traces caused by  protocols of different layers of Open Systems Interconnection (OSI) model. For example, the location information can be acquired through the GPS or IP address. (iii) \emph{Application data.} In addition, data originates from the use of third party services, e.g., playing online-games, which can be offered by either the same social service provides or other service providers \cite{Richthammer:ARES13}.
\item \emph{User-related Data} mainly refer to the data related to the personality and social interactions of users including the following sub-types: (i) \emph{User profile data} are profile-centric data which can describe personality aspects of users, e.g.,  address, education, favorites, hobbies, etc. (ii) \emph{Ratings/interests.} This type of data is mainly expressed interests of users, e.g., ``Likes'' and ratings of photos/posts shared by others. (iii) \emph{Social Network data.} One of the important features of social networks is \emph{small-world phenomenon}, which refers to the network-like relationships among people. The connections of social networks can  be either unidirectional or bidirectional. (iv) \emph{Contextual data.} There are some typical examples of contextual data including tagging peoples' names in their social networks, the status or the location of a shared item related to an event.
\end{enumerate}

\subsubsection{Internet of Things (IoT)}
\label{subsec:iot}


Internet of Things (IoT) can connect various \emph{things} to Internet so that data can be collected from the ambiance. The typical killer applications of IoT include the logistic management with Radio-Frequency Identification (RFID) technology \cite{ISO18000}, environmental monitoring with WSNs \cite{ZFei:CST2017}, smart homes \cite{stojkoska2017review}, e-health \cite{stankovic2017research}, smart grids \cite{yu2016smart}, Maritime Industry \cite{Hao:TENCON15}, smart city \cite{Mehmood:ComMag2017}, etc.

Wireless sensor networks (WSNs) can be regarded a sub-category of IoT technologies \cite{Ayaz:IEEESJ2018}. WSNs were first proposed to support military applications (e.g., surveillance in war zones) while WSNs have a wider range of applications rather than military surveilance. WSNs can be used in environment monitoring \cite{Boubrima:TWC2017,XBai:TCST2018}, ITS, smart manufacturing \cite{hndai:EIS19} and smart cities \cite{Memos:FGCS2018}. A sensor node usually consists of (i) a power module offering the reliable power, (ii) a sensor module gathering sensory data (via converting raw light, vibration and chemical signals into digital readings), (iii) a micro-controller processing the data received from the sensor and (iv) a wireless transceiver unit transferring the data to another sensor node or a sink. There are a number of wireless communication technologies proposed to support the data communications of WSNs including Bluetooth, IEEE 802.15.4 \cite{Raza:CST2017}, etc. Data sources of WSNs have similar features to the aforementioned networks including a wide range of data types (including temperature, light, pressure and speed), various physical dimensions and data heterogeneity. 

As another one of core technologies in IoT, RFID systems have been widely used in supply chain management, inventory control systems, retails, access control, libraries and e-health systems \cite{Ertek:TSMCS2017}. In particular, RFID allows a sensor (also called a reader) to read a unique identification  from a short distance without contacting with the tag \cite{WantR-PerCom06}. Data sources generated by RFID usually have the following characteristics: (a) RFID data contains noise and redundant information; (b) RFID data is temporal and streaming; (c) RFID data is processed on the fly; (d) RFID data volume is enormous.



Table \ref{tab:datasources} summarizes the key features of the aforementioned data sources. It is obvious that most data sources generate an enormous and heterogeneous data, which requires sophisticated data analytics. Therefore, we next discuss the necessities of BDA for large scale wireless networks.  

\subsection{Necessities of big data analytics for large scale wireless networks}
\label{subsec:necessity}
 
\begin{figure*}[t]
\centering
\includegraphics[width=15cm]{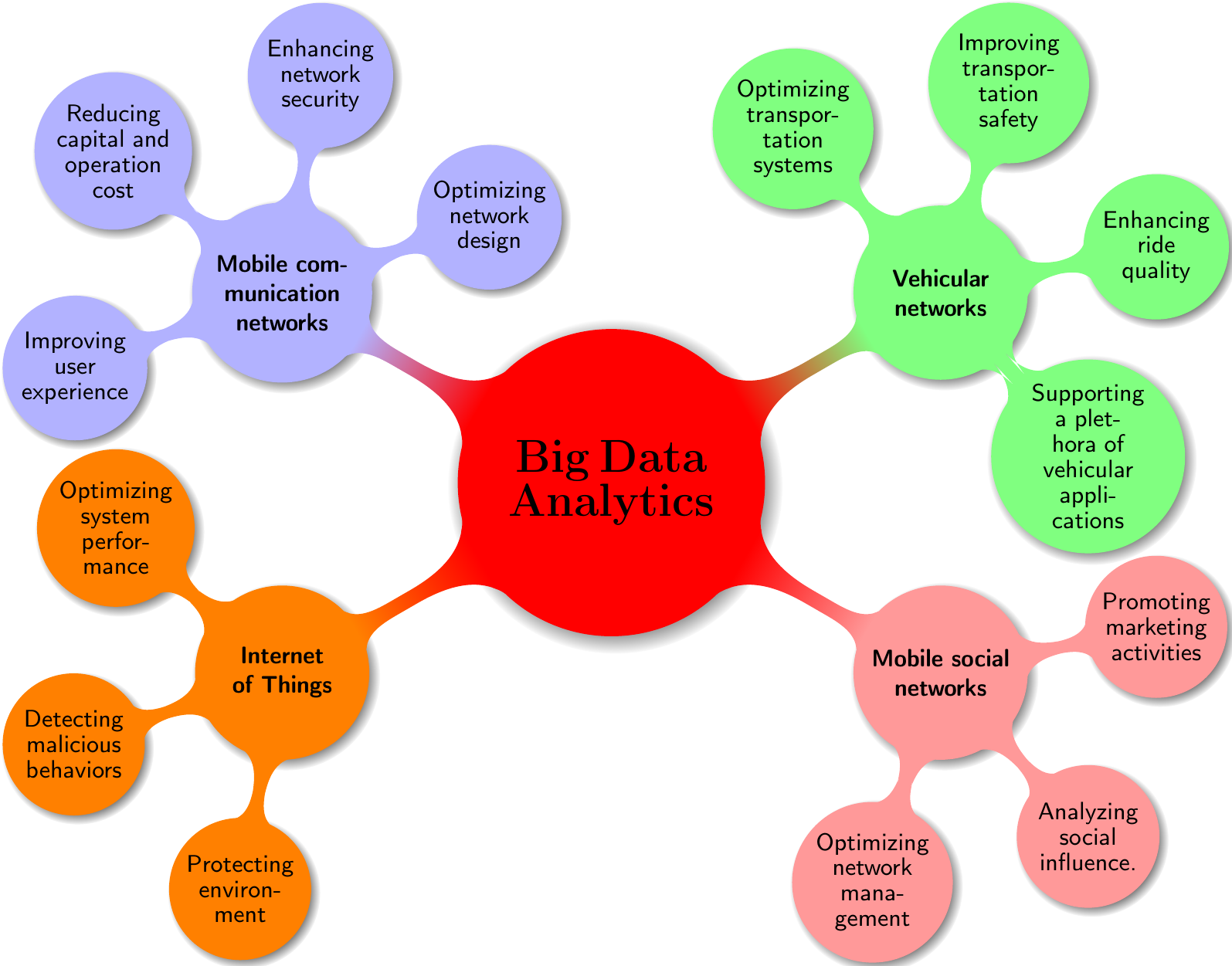}
\caption{Necessities of big data analytics for large scale wireless networks}
\label{fig:necessities}
\end{figure*}

There are an enormous amount of data generated everyday, which necessitates the demand that such ``big data'' need to be extensively analyzed so that some valuable and informative information can be obtained. In particular, we summarize the reasons of BDA for different types of wireless networks as shown in Fig. \ref{fig:necessities}.

\emph{1. Mobile communication networks}

One of the challenges that mobile network operators are facing is the growing cost (including capital expenditure and operating expenditure) and the energy consumption with the growth of user demands. It is necessary to optimize the cost and the energy consumption in mobile communication networks. The recent advances in data analytics have promoted the network optimization based on BDA \cite{CJiang:IEEEWC2017}. The benefits of BDA in mobile communication networks are summarized as follows.
\begin{itemize}
\item \emph{Improving user experience}. A mobile user always expects seamless network connectivity, omnipotent service, zero latency and low cost of service. BDA-based network optimization schemes can potentially improve the quality of user experience (QoE) and the quality of service (QoS) by analyzing both mobile service data and user data \cite{SBi:IEEEComMag15}.
\item \emph{Reducing capital and operational cost}. Mobile network operators (MNOs) can also benefit from BDA. First, MNOs can easily obtain both the data generated by users and the data generated by various network components. BDA can provide MNOs with deep insights before MNOs make the formal decisions \cite{KZheng:IEEENet16}. In particular, BDA technologies can extract intelligence and important information from various data, which can either be instantaneous or historical. The insightful information can help MNOs give both long-term strategies and make immediate decisions to reduce the capital and operational costs.
\item \emph{Enhancing network security}. BDA can also help to improve the network security. For example, the study of \cite{Sanctis:IEEENet2016} shows that using BDA can help to identify anomalies and malicious behaviors. Moreover, BDA can also be used in intrusion detection for next-generation networks \cite{gai2016intrusion}.
\item \emph{Optimizing network design}. BDA can help to optimize network design in both software-defined networks (SDN) \cite{LCui:IEEENet2016} and self-organizing network (SON) \cite{Mohajer:BDSON2017}. Moreover, it can be used to manage wireless traffic effectively \cite{ZLi:NSDI16}.
\end{itemize}

\emph{2. Vehicular networks}

The application of BDA in vehicular networks can bring a number of benefits as follows.

\begin{itemize}
\item \emph{Optimizing transportation systems.} BDA is the core of ITS. Specifically, BDA can provide ITS with important insights, such as planning public transit lanes, adjusting the length of traffic lights and predicting traffic flow \cite{POLSON:2017,ZZheng:IEEETIST19}. 

\item \emph{Improving transportation safety.} BDA also plays an important role in transportation safety. First, BDA can help to provide public with useful transportation information, such as traffic jams, road blockage due to an event and malfunctioning traffic infrastructures. Second, BDA also offers drivers real-time information about driving safety, such as collision avoidance, turn assistant and pedestrian crossing information \cite{Liu:TITS2017}.

\item \emph{Enhancing ride quality.} On one hand, BDA can offer drivers with the traffic situation and road information, through which drivers can plan a route with the minimum delay or can avoid traffic congestion. Moreover, BDA can help to analyze the user riding experience, consequently improving ride quality \cite{Furtado:BigData2017}.

\item \emph{Supporting a plethora of vehicular applications.} BDA can support a number of vehicular applications, such as weather and road information, interactive games and roadside services of nearby restaurants or gas-stations since most of them require the data from vehicular networks \cite{MacHardy:CST2018}. 

\end{itemize}

\emph{3. Mobile social networks}

With the revolution of web technologies and the proliferation of various mobile devices, enormous user data is generated from various mobile social services, including forums, blogs, microblogs, multimedia sharing services and wikis, though which people are virtually connected to form mobile social networks. BDA on mobile social networks can help us acquire valuable insights on personal interests, user behavior, social relations and concerns on media \cite{ZLv:TII2017}. In particular, BDA on MSNs can bring us a number of benefits in the following aspects.
\begin{itemize}
\item \emph{Promoting marketing activities.} Obtaining useful information of MSNs, we make better marketing decisions , promote product advertisements \cite{YBao:IWQoS2016} and enhance recommendation systems\cite{AMATO:FGCS2017}. 

\item \emph{Analyzing social influence.} BDA on MSNs can also help to predict political election \cite{STIEGLITZ:2018}, offer early warning of epidemics \cite{Atefeh:COIN15} and detect real-time events \cite{NGUYEN:FGCS2017}. Moreover, it can be used to detect anomalous behaviors in social networks \cite{ruan2016profiling}. 

\item \emph{Optimizing network management.} BDA on MSNs is beneficial to the network optimization \cite{ZSu:IEEENet16} and the location-based service design \cite{Hristova:WWW2016}. Furthermore, it can be used to increase the routing efficiency and reliability in mobile ad hoc networks \cite{ZhangSong:ComMag2017}.
\end{itemize}

\emph{4. Internet of Things}


In particular, BDA on IoT can bring us a number of benefits in the following aspects.
\begin{itemize}
\item \emph{Protecting environment.} Sensors mounted in IoT can collect various ambient data, which can then be used to identify possible environmental hazards and offer decision support on environmental protecting policies\cite{Montori:IoTJ2018}. Moreover, the real-time analysis on industrial environmental data can also help to make an immediate response to emergencies \cite{GZhu:IJSDWE2017}. 

\item \emph{Detecting malicious behaviors.} The analysis on massive data in IoT can be used to detect malicious behaviors. For example, it is shown in \cite{ZZheng:TII18} that the analysis on smart grid data can help to detect  electricity theft consequently securing smart grids. Moreover, the IoT data in the whole food supply-chain is also beneficial to prevent mischievous actions and guarantee food safety \cite{Leng2018}.

\item \emph{Optimizing system performance.} The data analysis on the IoT-enabled supply chain can help to improve the system efficiency and reduce the turnaround time \cite{Dweekat:IMDS2017}. In addition, BDA on IoT-enabled intelligent manufacturing shops \cite{RayZhong:IJPR2017} can also help to make accurate logistic plan and schedules. As a result, the system efficiency can be greatly improved.
\end{itemize}

\section{Data Acquisition}
\label{sec:acquisition}

We first discuss the challenges in Section \ref{subsec:chall_data_acq}. We then introduce current solutions to these challenges in four types of wireless networks in Section \ref{subsec:data_studies_acq}. We finally discuss research opportunities in Section \ref{subsec:sum_data_acq}.

\subsection{Challenges in data acquisition}
\label{subsec:chall_data_acq}

There are the following challenges in data acquisition (including data collection and data transmission).
\begin{itemize}

\item \emph{Difficulty in data representation}. Due to the high diversity of data sources, the data sets in large scale wireless networks have different types, heterogeneous structures and various dimensions. Take mobile communication networks as an example. There are both user data (such as voice, text and video) and system data (including cell-level and core-network-level data). How to represent these structured, semi-structured and un-structured data becomes one of major challenges in BDA for large scale wireless networks.

\item \emph{Effective data collection}. Data collection refers to the procedure of obtaining raw data from various types of wireless networks. This process must be effective and valid since the inaccurate data collection will affect the subsequent data analysis procedure. 

\item \emph{Efficient data transmission}. How to transmit the tremendous volumes of data to data storage infrastructure in an efficient way becomes a challenge due to the following reasons: (i) \emph{high bandwidth consumption} since the transmission of big data becomes a major bottleneck of wireless systems; (ii) \emph{energy efficiency} is one of major constraints in many wireless systems, such as wireless sensor networks and IoT.

\end{itemize}

We then present current solutions in data acquisition in large scale wireless networks.

\subsection{Current solutions in data acquisition}
\label{subsec:data_studies_acq}

\subsubsection{Mobile communication networks (MCNs)} 
\label{subsec:mcn-da}

\begin{figure}[t]
\centering
\includegraphics[width=8.5cm]{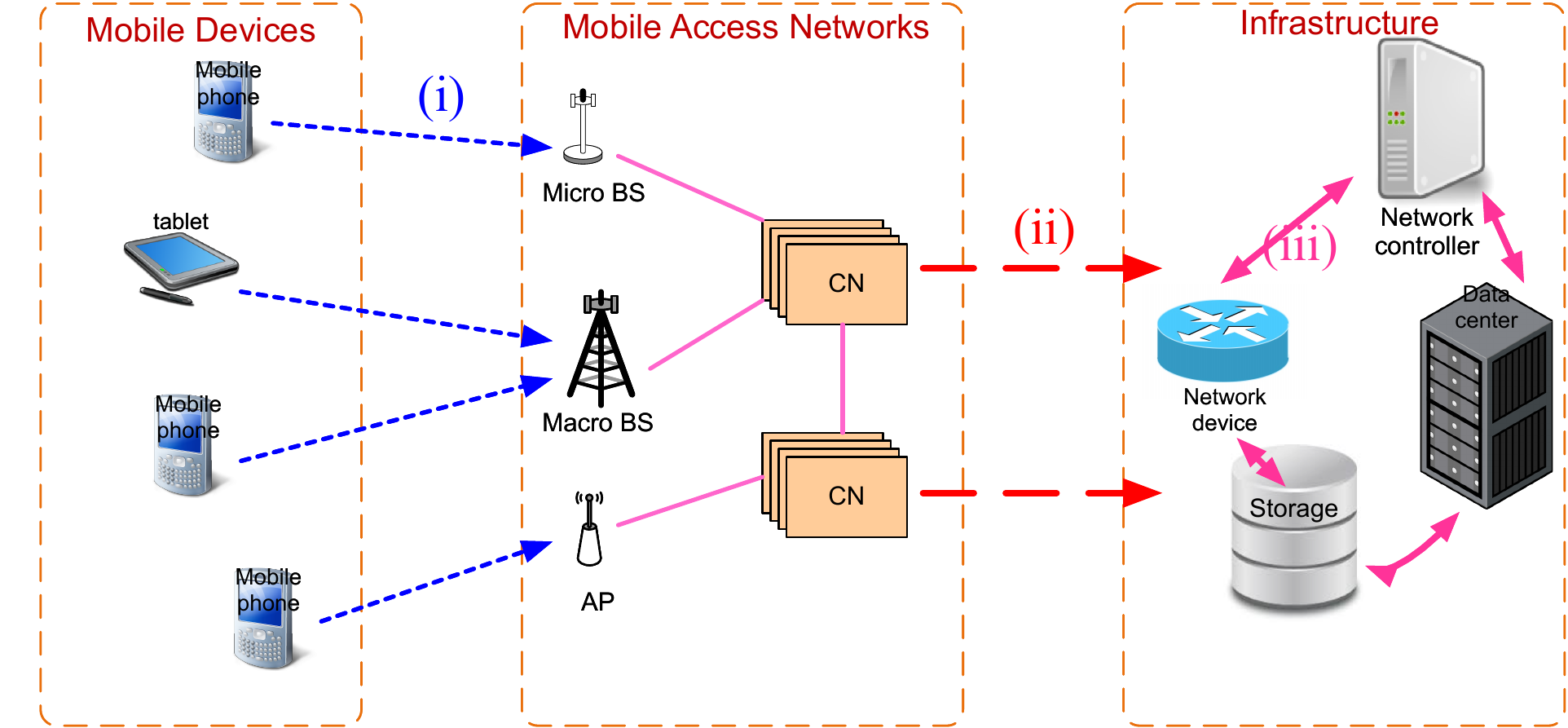}
\caption{Data transmission in mobile communication networks}
\label{fig:mobilenet}
\end{figure}
It is a critical issue to represent various types of data in MCNs. In order to support further data preprocessing and analytics, radio waveform data and signaling data are usually converted and represented in matrices or vectors \cite{YHe:IEEEAccess2016} while call detail records (CDRs) and user profiles are represented in records in DBMS \cite{AImran:IEEENet14}. User data is stored at mobile devices, base stations and processing units while system data is usually stored at base stations and processing units. 

Both user data and system data in cellular networks are mainly collected in the \emph{pull-based} manner \cite{Biral:DCN15}, in which data is collected \emph{proactively} by agents distributed in the whole network. In conventional cellular networks, system data is usually stored at some centralized servers offered by mobile operators. However, the proliferation of massive data in mobile networks leads to the big challenge in managing both user data and system data in heterogeneous networks. Content-centric networks are one of possible solutions to this challenge \cite{Su:IEEEComMag15}. The main idea of content-centric networks is to cache the popular contests at the intermediate servers (base stations, gateways or routers) so that the user demands for the same content can be fulfilled locally \cite{Wang:IEEEComMag14}. As a result, a lot of traffic can be significantly reduced.

The collected data will be further transmitted to the storage infrastructure. As shown in Fig. \ref{fig:mobilenet}, data transmission procedure of wireless networks typically consists of the following sub-phases \cite{SBi:IEEEComMag15}: (i) transmission from mobile devices to base stations (or APs), which are connected with core networks (CNs); (ii) transmission from CNs to storage infrastructure and computing infrastructure (i.e., data centers); (iii) transmission between data centers. In sub-phase (i), the communications are mainly conducted through wireless connections, which have the \emph{limited capacity}, are vulnerable to interference and are susceptible to eavesdroppers compared with conventional wired networks. Sub-phase (ii) mainly consists of wired links connecting base stations to CNs and the links connecting CNs to data centers. Similar to Sub-phase (ii), Sub-phase (iii) consists of wired links, which usually have the higher bandwidth and are more robust than wireless links \cite{Bhaumik:mobicom2012}. 

The connections between BSs and CUs is named as \emph{front-haul} links \cite{Checko:IEEECST15}. CUs are connected with mobile core networks (CNs) through \emph{back-haul} links. It is challenging to manage network resources in both front-haul and back-haul networks effectively in order to support big data applications in next generation mobile networks \cite{SBi:IEEEComMag15}. C-RAN \cite{Checko:IEEECST15} is one of the most promising architectures with cost-efficient and energy-efficient solutions. SON \cite{Mohajer:BDSON2017}, HetNets \cite{Mehmeti:TMC2017} and software defined networking (SDN) \cite{Fan:IEEENet16,Kuang:IEEENet16} are other proposals.


\subsubsection{Vehicular networks (VNets)}


\begin{figure}[t]
\centering
\includegraphics[width=8.5cm]{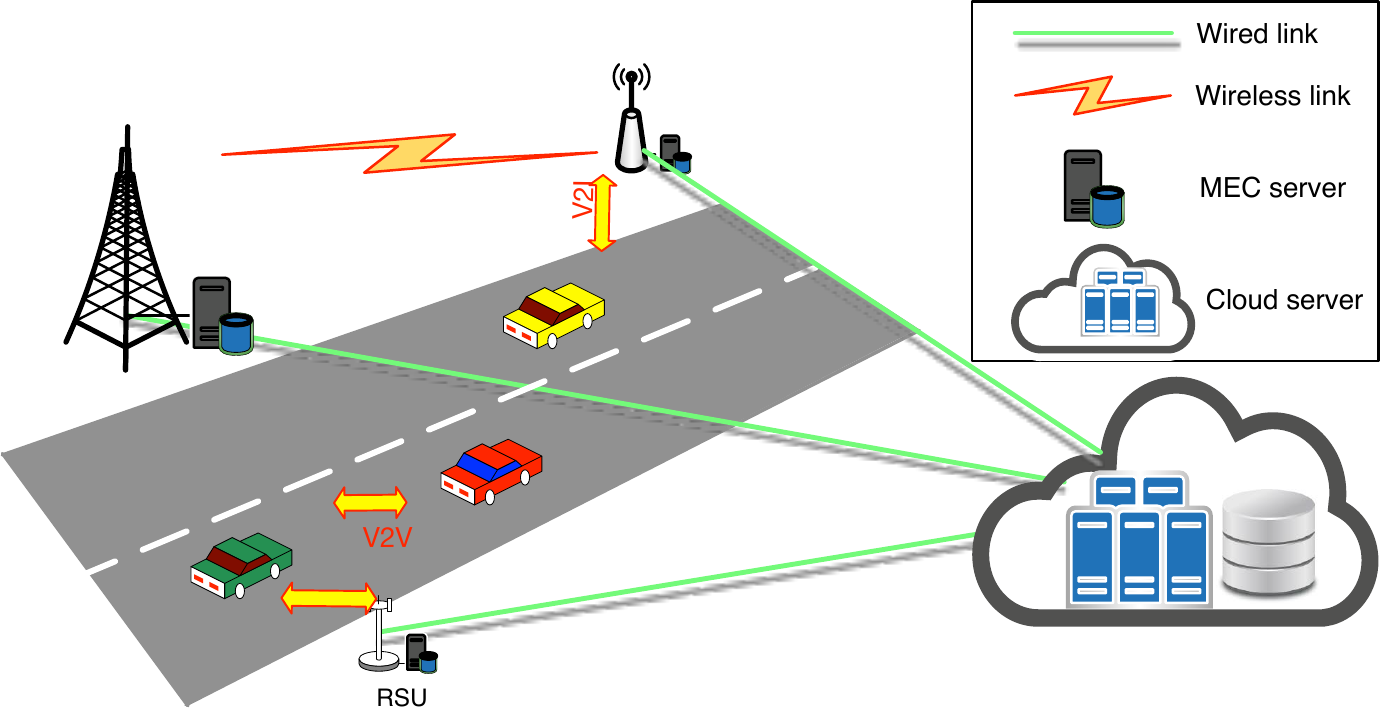}
\caption{Data acquisition in vehicular networks}
\label{fig:vnets}
\end{figure}

Fig. \ref{fig:vnets} shows different types of vehicular communications to support data acquisition. For example, data can be transmitted through vehicle to vehicle (V2V), vehicle to road side unit (V2R), vehicle to infrastructure (V2I) manners. The vehicular data can be collected and preprocessed at Mobile Edge Computing (MEC) servers located at RSUs, BSs or at remote clouds. However, data collection of VNets is suffering from the rapid changed network topology due to the movement of vehicles. How to design delay-tolerant data gathering schemes in distributed VNets has received extensive attention recently. In \cite{Mansour:2015}, a new data gathering and distribution scheme named Collaborative Data Collection Protocol (CDCP) was proposed. The main feature of CDCP lies in the optimized delay and it can consequently be used in non-delay tolerant applications. In \cite{Brik:2016}, a novel Distributed Data Gathering Protocol (DDGP) was proposed for data collection conducted by vehicles in highways. The main idea of DDGP is to allow vehicles to access the channel in a distributed manner according to their geo-locations. Besides, DDGP removed those redundant, dated and undesired data. As a result, DDGP improves both the reliability and the efficiency of the data collection process compared with other existing schemes. In addition, data collected in VNets are usually represented in data records or text logs\cite{Ilarri:CST2015}, which will be further preprocessed and stored at vehicles, RSUs, BSs and at ITS centers.

How to reduce the amount of data is another challenge in data acquisition of VNets. In \cite{placzek2017efficient}, a data transmission scheme based on the estimation of data flow at control nodes. With the support of traffic flow estimation and uncertainty estimation, the amount of transmitted data can be greatly reduced. The study of \cite{Sahoo:ITS2017} proposed a hierarchical aggregation scheme to reduce the amount of data to be transferred in VNets. Besides, data compression schemes \cite{Gandhi:sigmod2009} used in WSNs can also be applied to VNets.

\subsubsection{Mobile Social Networks (MSNs)}


\begin{figure}[h]
\centering
\includegraphics[width=8.5cm]{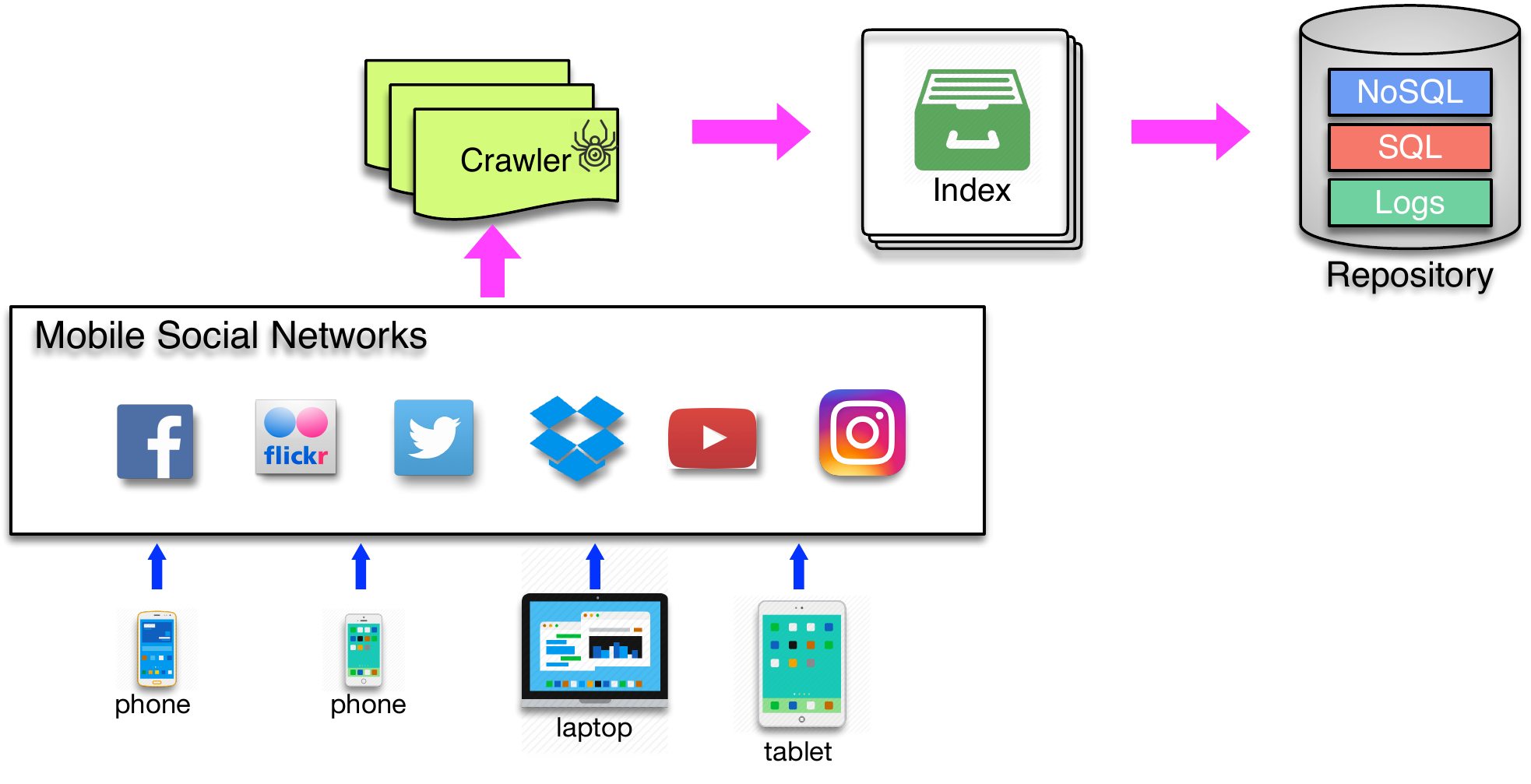}
\caption{Data acquisition in mobile social networks}
\label{fig:MSN}
\end{figure}

Conventional data gathering methods used in web technologies such as web crawlers to social networks can be used to collect the data from mobile social networks. Fig. \ref{fig:MSN} shows an example of using web crawlers to obtain mobile social data. Besides web crawlers, log files stored at web servers also gather useful information, such as the clicks, hits, access time and other important attributes, which are either represented in log files or non-SQL data format (like json) \cite{leskovec2016snap} \cite{HAJARIAN2017}. In addition to web crawlers and log files, we can use various sensors to acquire user-related data from mobile devices. This emerging technology namely mobile crowdsensing (MCS) has received extensive attention recently. There are many challenges in MCS such as coverage constraint \cite{GHan:ComMag17}, incentive mechanisms \cite{GYang:ComMag17}, privacy preservation \cite{Alsheikh:ComMag17} and energy consumption \cite{JWang:ComMag2018}. 

The collected data from mobile social networks will then be sent through mobile communication networks to mobile social service providers for the further analysis. Similar to mobile communication networks, mobile social services providers also have data centers to store the massive data from mobile social networks while the above research challenges should be addressed. Since the data transmission follows the procedure similar to that of mobile communication networks, we omit the discussions here.

\subsubsection{Internet of Things} 
\label{subsec:wsn-da}

\begin{table}[h]
\caption{Comparison between data-on-network tags and data-on-tag tags}
\centering
\footnotesize
\renewcommand{\arraystretch}{1.75}
\begin{tabular}{|m{1.4cm}|m{3cm}|m{4cm}|}
\hline
& \textbf{DON tags} & \textbf{DOT tags} \\
\hline
\hline
Information & ID & object related data, such as time, location, temperature, etc.\\
\hline
Storage & low storage capacity & high storage capacity\\
\hline
Energy Supply & No & Yes (some) \\
\hline 
Data Access & network connection & presence of objects \\
\hline
Security & Access control at infrastructure & encryption at tags\\
\hline
Cost & low & high\\
\hline
\end{tabular}
\label{tab:rfid-compare}
\end{table}

RFID tags allow the uniquely identifiers attaced at objects to be read in a short distance by a RFID reader in wireless manner. RFIDs tags can be categorized as data-on-network (DON) and data-on-tag (DOT) types \cite{Diekmann:2007}. Table \ref{tab:rfid-compare} compares the two different types of RFIDs. In particular, Electronic Product Code (EPC) tag is one of the typical DON RFIDs. In an EPC tag, there is no extra storage except for the product ID information. Usually, an EPC tag can be passively read by an RFID reader periodically and there is no onboard battery on an EPC tag. This design can greatly save the cost. Due to the unstable channel conditions such as the shiedings and reflections \cite{Kennedy:BigData2017}, the EPC reads may contain errors and noise. Therefore, filters are needed to preprocess the raw EPC reads. Then, readers aggregate the preprocessed data into events, which are stored in EPC systems and can be further accessed by other EPC applications. With the decreased cost of RFID transponders, DOT types of RFIDs will become affordable in the future. Compared with DON tags, DOT tags have the higher storage and can be supplied by onboard batteries. Besides, some sensors (e.g., temperature sensors) can be mounted with DOT tags that can store the sensor information (such as temperature and humidity) \cite{WantR-PerCom06}. Some of these DOT tags can actively transmit the information to other nodes. 

\begin{figure}[t]
\centering
\includegraphics[width=8.5cm]{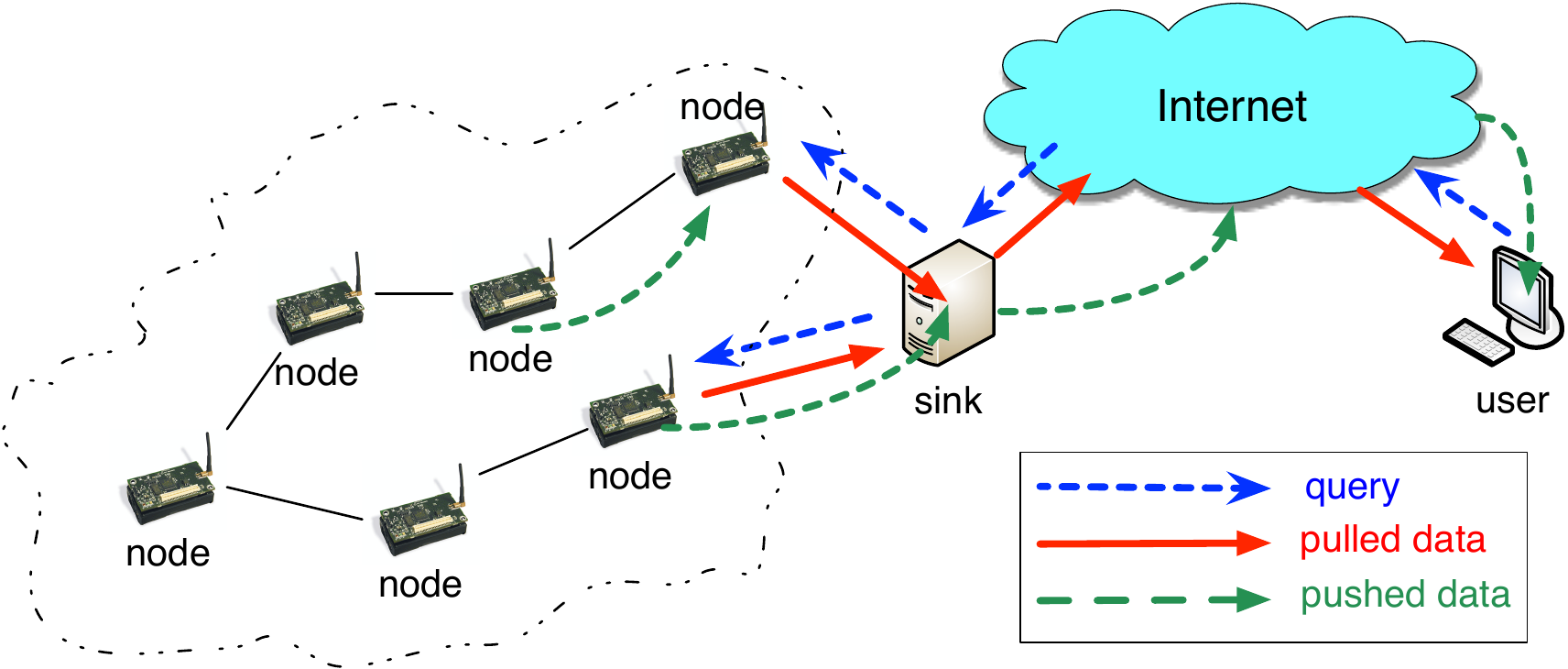}
\caption{Pull-based methods vs Push-based methods}
\label{fig:WSN-pushpull}
\end{figure}

Another key IoT technology is WSNs. There are two major approaches for data collection in WSNs: \emph{pull-based} methods and \emph{push-based} methods \cite{fahmy2016wireless}. Fig. \ref{fig:WSN-pushpull} compares the two methods. In the pull-based schemes, users first send queries to the sink, which then broadcasts the queries to sensor nodes. Then, the data is sent to the sink according the specified queries and the sink next forwards the data to the users. In the push-based schemes, the sensors autonomously decide when to send the sensor data to the sink that consequently forwards the data to the users. 

One of major challenges in data transmission in WSNs is the energy consumption since most of sensor nodes have the limited energy (i.e., supplied by batteries). How to design energy-efficient routing/transmission schemes in WSNs has received extensive attention recently. The study of \cite{ZChen:MIS2017} proposed an energy-efficient one-to-many broadcasting scheme for data collection in WSNs. Ref. \cite{Abdul-Salaam:IEEESensor2017} developed an energy-efficient data collection scheme for position-free WSNs. In addition to energy consumption, the transmission delay has also been considered in recent works \cite{Takaishi:IEEETETC14,SYang:TMC2017}.

Conventional RFID and WSNs that are suffering from short communication range (typically less than hundreds of meters) cannot support the wide-coverage scenarios like smart metering, smart cities and smart grids \cite{JXu:IOTJ2017}. Low Power Wide Area (LPWA) networks essentially provide a solution to the wide coverage demand. Typically LPWA technologies include Sigfox, LoRa, Narrowband IoT (NB-IoT) \cite{MEKKI2018}. One of LPWA advantages is low power consumption. For example, NB-IoT has a ten-year battery life \cite{JXu:IOTJ2017}. Moreover, LPWA has a longer communication range than RFID and WSNs. Specifically, LPWA technologies have the communication range from 1km to 10 km. Moreover, they can also support a large number of concurrent connections (e.g., NB-IoT can support 52,547 connections \cite{JXu:IOTJ2017}). However, one of limitations of LPWA technologies is the low data rate (e.g., NB-IoT can only support a data rate upto 200 kps). Therefore, LPWA technologies shall complement with conventional RFID and WSNs so that they can support the various data acquisition requirements.

In addition to LPWA technologies, Wireless LAN is another alternative  solution to data acquisition in IoT. The recent work in \cite{Iqbal:PIMRC18} presents a data acquisition scheme based on IEEE 802.11AH standard to ensure reliable seismic data acquisition.


\subsection{Opportunities in data acquisition}
\label{subsec:sum_data_acq}


Although the challenging issues as mentioned in Section \ref{subsec:chall_data_acq} have been partially or fully addressed, it is worth mentioning that there are still many issues not well addressed. We identify three research opportunities as follows:

\begin{itemize}
\item \emph{Heterogeneity of data types.} There are different types of data in each type of wireless networks. For example, user data in mobile communication networks is usually unstructured or semi-structured in contrast to structured operator data. There is no general representation method or tool to depict the heterogeneous types of data. This may result in difficulties in data preprocessing, data storage and data analytics. Therefore, research on handling heterogeneous data types will be a new trend in this area.

\item \emph{Security in data transmission.} Data transmission in IoT, RFID and WSNs is often vulnerable to malicious attacks due to the limitation of IoT objects, RFID tags and sensor nodes. For example, the security module of narrow-band IoT is removed to save the cost of NB-IoT objects \cite{JXu:IOTJ2017}, which however makes the susceptibility of IoT objects to security threats. Recent research efforts like secure key generations based on reciprocity and randomness of wireless channels \cite{WHu:IoTJ19} and protective jamming schemes \cite{LHu:IoTJ18} have shown the effectiveness in IoT.

\item \emph{Energy harvesting in data transmission.} In addition to the security issues during the data transmission in IoT, how to provide the low-power RFIDs or other objects in IoT with the energy is another challenge. Recently, RF-enable wireless energy transfer technology \cite{BiHoZhang:IEEEComMag15} provides an attractive solution by powering the RFID tags with energy over the air. However, in order to overcome the higher attenuation of RF energy, directional wireless power transfer is expected \cite{ZWang:2016}. 
\end{itemize}

\section{Data Preprocessing}
\label{sec:preprocessing}

We first discuss the challenges in Section \ref{subsec:chall_preprocessing}. We then introduce exiting studies in data acquisition in four types of wireless networks in Section \ref{subsec:data_preprocessing}. We discuss the research opportunities in Section \ref{subsec:summary_datapreprocessing}.

\begin{figure*}[t]
\centering
\includegraphics[width=16.5cm]{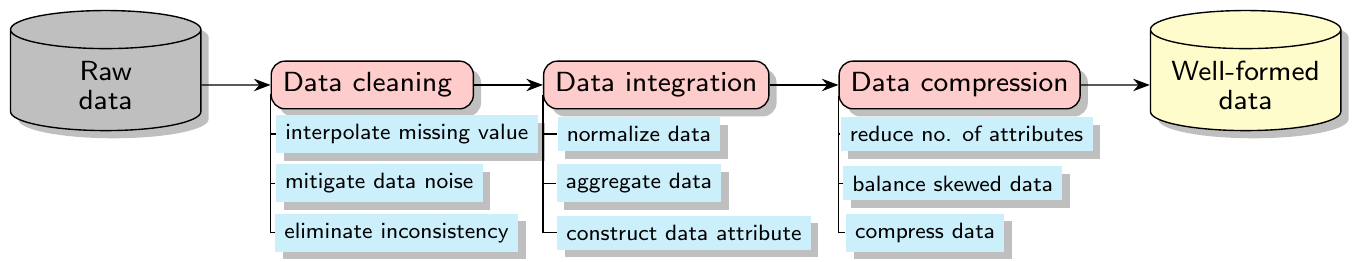}
\caption{Data preprocessing procedure}
\label{fig:preprocessing}
\end{figure*}

\subsection{Challenges in data preprocessing}
\label{subsec:chall_preprocessing}

Data acquired from large scale wireless networks has the following characteristics:

\begin{itemize}

\item \emph{Various data types}. There are various data types generated from large scale wireless networks, including text, sensed values, audio, video, etc. The data is structured, semi-structured and non-structured. 

\item \emph{Erroneous and noisy data}. The data obtained from wireless networks are often erroneous and noisy mainly due to the following reasons: (a) intermittent loss of communications, (b) the failure of wireless nodes or sensors and (c) interference during the process of data collection \cite{MLi:TON2009}. For example, wireless communications are often interfered by various channel conditions, such as blockage, fading and shadowing effects. Besides, in wireless sensor networks, the data collection may fail when sensors deplete their batteries. 

\item \emph{Data duplication}. Data generated in large scale wireless networks often contain excessive duplicated information, exhibiting in both temporal and spatial dimensions. For example, the enormous duplicated RFID data will be obtained when several readers read multiple RFID tags at different time moments \cite{Ertek:TSMCS2017}; this data duplication often results in data inconsistency. Besides, as shown in wireless health-care systems \cite{YZhang:IEEESysJ2017}, a vast volume of duplicated medical data has been generated in real-time fashion. It is worth mentioning that data redundancy is often beneficial in DBMS in order to improve the data reliability while data duplication results in data inconsistency especially in data preprocessing.

\end{itemize}

The above features lead to the following research challenges in data preprocessing.

\begin{itemize}

\item \emph{Integration of various types of data}. As mentioned above, data generated in large scale wireless networks has the various types and heterogeneous features. It is necessary to integrate the various types of data so that efficient BDA schemes can be implemented. However, it is quite challenging to integrate various categories of data. 

\item \emph{Duplication reduction}. An aforementioned challenge lies in the temporal and spatial duplication of the raw data generated in wireless networks. The data duplication often leads to the data inconsistency, which has the impacts on the subsequent data analysis. 

\item \emph{Data cleaning and data compression}. In addition to data duplication, data of large scale wireless networks is often erroneous and noisy; it inevitably makes data cleaning more difficult. Thus, we need to design effective schemes to compress data and clean the errors of data.  

\end{itemize}

We next present existing studies (i.e., solutions) in data preprocessing in BDA for large scale wireless networks.

\subsection{Existing studies in data preprocessing}
\label{subsec:data_preprocessing}


Fig. \ref{fig:preprocessing} shows the three major steps in converting raw collected data into well-formed data: 1) data cleaning, 2) data integration and 3) data compression, each of which consists of different preprocessing techniques.

\subsubsection{Mobile communication networks} 
\label{subsec:mcn-dp}


The typical data preprocessing techniques include \emph{data integration}, \emph{data cleaning} and \emph{data duplication elimination}. Data integration approaches typically include data warehouse method \cite{Lenzerini:2002} and data federation method \cite{Hass:IBM02}. However, both the two methods may not be applicable to massive data generated from heterogeneous networks. In this sense, we may exploit data integration schemes dedicated for massive data \cite{Gubanov:ICDE2017} of mobile communication networks. In order to remove the erroneous, inaccurate, incomplete data, data cleaning is necessary. The typical data cleaning schemes (models) include regression models, probabilistic models and outlier detection models \cite{Han:2012}. Data duplication is a common issue in big data of wireless networks. The typical solutions include duplication detection \cite{Zhang:SIGIR2002} and data compression \cite{Gandhi:sigmod2009}. In addition to the above data preprocessing schemes, other approaches such as dimensionality reduction and feature selection are also useful in data preprocessing \cite{Sanctis:IEEENet2016,Khatib:ComMag16}. Moreover, a data cleaning model has been proposed in \cite{Fan:TKDE16} to understand user preference.

\subsubsection{Vehicular networks}

In vehicular networks, vehicles and other units such as Road Side Unit (RSU) produce a tremendous amount of data. Data compression is one of typical strategies to reduce the volume of the data. In particular, Feldman et al. in \cite{Feldman:IPSN2012} proposed a method named Core-Set to compress the streaming data generated from distributed vehicular networks. The main idea of Core-Set is to construct a small set that approximately represents the original data. Core-Set can also be used to compress the diversity of massive data sets generated from in-car GPS and cameras. Moreover, a cooperative data sensing and compression approach was proposed in \cite{XYu:ICC10} to compress the data and reduce the communication traffic. This method is mainly based on exploiting the spatial correlation of  the data.

In addition to data compression, data of vehicular networks can also contain many errors or incompleted data instances. Therefore, we can apply the aforementioned data cleaning schemes such as regression models, probabilistic models, outlier detection models \cite{Han:2012} to remove the errors and the duplicated data. For instance, it is shown in \cite{Fogue:TMC14} that there is no noise or inaccuracies detected in data of automotive accidents after applying the above data cleaning and other data mining approaches. 
 
\subsubsection{Mobile Social Networks}

Data preprocessing schemes of mobile social networks include data integration, cleaning, and duplication elimination. Similarly, we can use the typical data warehouse method and data federation method to integrate the service-provider-related data of mobile social networks. However, it is quite challenging to integrate the {\it user-related data} since the above conventional methods cannot be used to integrate mobile sensing data and online social network (OSN) data \cite{Mehrotra:middleware2014}. Besides, the privacy preservation during the data integration is also necessary \cite{Aggarwal:2011}. Moreover, the missing values can be restored by data augmentation  \cite{JORGENSEN:2018}.

Data duplication is a critical issue in mobile social networks. There are several proposals in addressing the above challenges. In \cite{Zheng:www2010}, the integration of local information (obtained from GPS) and activity recommendations was investigated. In particular, only valuable information is extracted. Moreover, Mehrotra et al. \cite{Mehrotra:middleware2014} proposed and implemented a middleware named SenSocial, which can obtain mobile sensory data and integrate with OSN automatically. The two implemented prototypes demonstrated that SenSocial can significantly reduce the efforts in developing OSN applications.

\subsubsection{Internet of Things}

RFID data is generally noisy, erroneous and redundant because of complicated wireless channel conditions and cross-reads from multiple readers \cite{Ertek:TSMCS2017}. Hence, it is necessary to preprocess the raw RFID data. Data preprocessing approaches on RFID data include \emph{data cleaning} and \emph{data compression}. In \cite{Baba:2017}, an Indoor RFID Multi-variate Hidden Markov Model (IR-MHMM) was proposed to identify data uncertainty and clean the cross-reads of the RFID data. RFID data also contains valid reads that nevertheless is non-of-interest for analysis. The study of \cite{MA:ESA2018} proposed a machine-learning based method to filter out this useless data.

In WSNs, sensor data is usually uncertain and erroneous due to the depletion of battery power, imprecise measurement of sensors and network failures. To address these issues, it is necessary to employ data cleaning schemes. However, data cleaning in sensor data is challenging since there are strong temporal and spatial correlations between sensor data. There are several approaches proposed to address this challenge. In particular, an autocorrelation-based scheme was propsed in \cite{Bhandari:sensors17} to preprocess time-series temperature data and remove duplicated data. In \cite{Tasnim:CCNC17}, a novel data cleaning mechanism was proposed to clean erroneous data in environmental sensing applications in WSNs. In WSNs, energy-saving is a critical issue in data-cleaning algorithms. In \cite{CDeng:IJPR2018}, an energy-efficient data-cleaning scheme was proposed. In addition, an interpolation method was proposed in \cite{ZZheng:TII18} to recover the missing values of smart grids. Moreover, the work of \cite{Aleman:ICNC18} presents a context-aware data cleaning scheme to estimate and restore the missing values of WSNs while minimizing the error.

\subsection{Opportunities in data preprocessing}
\label{subsec:summary_datapreprocessing}


Although most of the challenges as mentioned in Section \ref{subsec:chall_preprocessing} have been solved, it is worth mentioning that there are still many issues not well addressed. We just enumerate some of research opportunities as follows:

\begin{itemize}
\item \emph{Privacy preservation in data preprocessing.} Most of existing studies preprocess data without consideration of the protection of sensitive information of users. For example, in order to improve user experience in text input in mobile devices, mobile operating systems often collect the frequently-used terms from users and preprocess them \cite{xkxiao:icde2018}. It however would violate user's privacy. How to design privacy-preserved data preprocessing schemes becomes a critical issue.

\item \emph{Security assurance in data preprocessing.} Data preprocessing has been conducted in different sub-systems throughout large scale wireless networks. The fragmentation and heterogeneity of wireless networks nevertheless often result in the difficulty in guaranteeing the security during data preprocessing. For example, it is shown in \cite{Roman:CN2013} that IoT systems are also vulnerable to be malicious attacks due to the failure of security firmware updates in time. Although typical solutions such as authentication, authorization and communication encryption can partially amend security vulnerability, the general solutions across heterogeneous wireless systems are still expected. 

\item \emph{Energy-efficiency in data preprocessing.} RFID tags and wireless sensors are suffering from the limited energy. Therefore, heavy-weighted data preprocessing schemes may not be feasible at these IoT nodes. Many existing studies only consider uploading raw data to remote clouds, which preprocess the data. However, it may cause extra delay to upload and process data at the clouds. Mobile Edge Computing (MEC) can help to offload processing tasks at local MEC serves so that the delay can be greatly reduced.
\end{itemize}

\section{Data Storage}
\label{sec:storage}

\begin{figure*}[t]
\centering
\includegraphics[width=16.5cm]{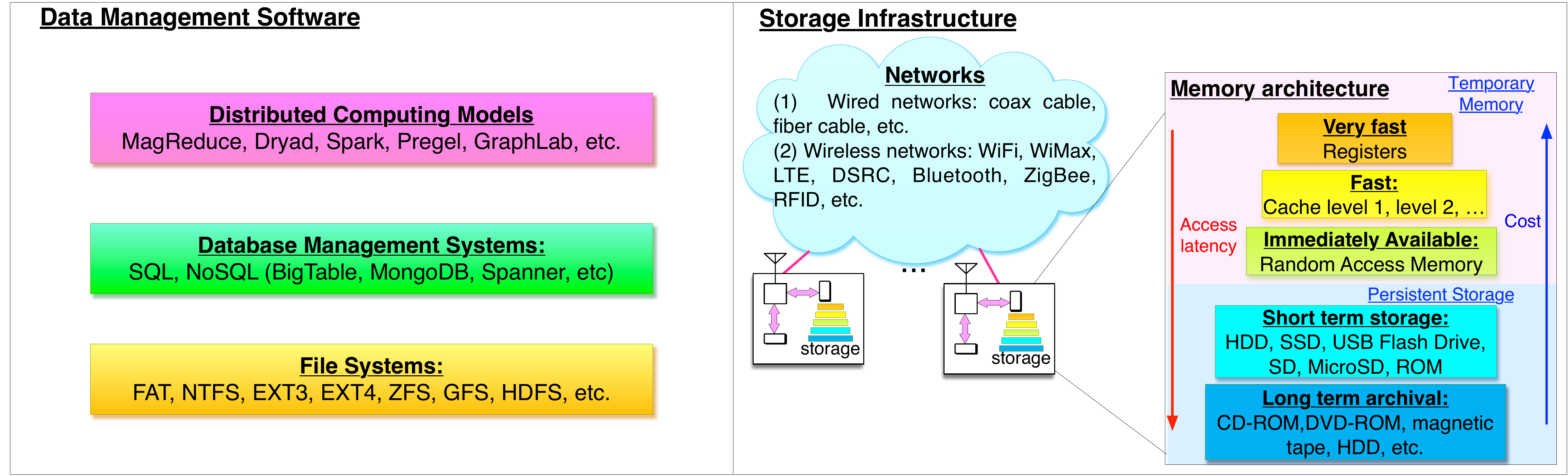}
\caption{Data storage system}
\label{fig:storage}
\end{figure*}

Data storage plays an important role in big data analytics for large scale wireless networks. We first summarize the challenges of data storage in Section \ref{subsec:challenges-storage}. Fig. \ref{fig:storage} illustrates a general data storage system used for big data analytics for large scale wireless networks. In particular, the data storage architecture can be categorized into two layers: storage infrastructure to be introduced in Section \ref{subsec:infrastructure} and data management software to be presented in Section \ref{subsec:dataman}. We finally discuss the research opportunities in Section \ref{subsec:opp-storage}.

\subsection{Challenges in data storage}
\label{subsec:challenges-storage}

Data storage plays an important role in data analysis. However, designing an efficient and scalable data storage system is challenging in large scale wireless networks. We summarize the challenges in data storage as follows.

\begin{itemize}

\item \emph{Reliability and persistency of data storage}. Data storage systems must ensure the reliability and the persistency of data. However, it is challenging to fulfill the above requirements of big data systems while balancing the cost due to the tremendous amount of data \cite{Guerra:FAST2011}. 

\item \emph{Scalability}. Besides the storage reliability, another challenging issue lies the scalability of storage systems for BDA. The various data types, the heterogeneous structures and the large volume of massive data sets of wireless networks lead to conventional databases being infeasible in BDA. 

\item \emph{Efficiency}. Another concern with data storage systems is the efficiency. In order to support the vast number of concurrent accesses or queries from the data analytics phase, data storage needs to fulfill the efficiency, the reliability and the scalability together, which is extremely challenging.

\end{itemize}

We next summarize the solutions to the above challenges in this section; we categorize these solutions according to storage infrastructure and data management software (as shown in Fig. \ref{fig:storage}) to be presented in  Section \ref{subsec:infrastructure} and Section \ref{subsec:dataman}, respectively.

\subsection{Storage Infrastructure}
\label{subsec:infrastructure}

Storage devices can be categorized into the following types according to the storage methods \cite{Placek:rep2006,Goda:PIEEE12}:

\begin{enumerate}
\item \emph{Persistent Storage} devices mainly include: i. long term storage (for archival usage): magnetic Harddisk Drives (HDDs), magnetic taps, CD-ROMs, DVD-ROMs, etc.; ii. short term storage: HDDs, Solid-State Drives (SSD), USB flash drives, Secure Digital (SD) cards, micro SD cards, Read-Only-Memory (ROM), etc.

\item \emph{Temporary Memory} devices mainly include: i. immediately available memory: Random Access Memory (RAM); ii. Fast memory: Caches (level 1, level 2, etc.) at CPUs or other processing units; iii. Very fast memory: registers at CPUs or other processing units.

\end{enumerate}

The memory architecture shown in Fig. \ref{fig:storage} also indicates the different performance metrics and the cost of different storage devices. In particular, persistent storage devices usually have the lower cost than temporary memory devices while the access latency of persistent storage devices is also significantly higher than that of temporary memory. For example, HDDs are usually 100,000 slower than registers inside CPUs \cite{Patterson:2013}. To balance the cost and the access latency of heterogeneous storage devices, multi-tier hybrid storage architectures have been proposed recently in \cite{Guerra:FAST2011,Strunk:2012,Cheng:hpdc2015,Li:tos2015}.

There are various wireless devices including smart-phones, tablets, PCs, laptops, vehicles, GPS navigators, sensors, RFID tags, IoT objects and wearable devices, etc. Due to the heterogeneity of wireless devices, each wireless device may only include subsets of the aforementioned storage devices but not all of them. For example, a smart-phone may consist of the persistent storage devices, such as ROM and a micro SD card, as well as the temporary memory devices, such as RAM and registers while there is no HDD in a smart-phone due to the bulky size of HDDs. However, an EPC RFID tag may only contain a Complementary metal-oxide-semiconductor (CMOS) integrated circuits with Electrically Erasable Programmable Read-Only Memory (EEPROM) with limited storage capacity \cite{Landt:IEEEPot05}. 

\begin{table*}[t]
\caption{Comparison between SQL Database and NoSQL Database}
\centering
\footnotesize
\renewcommand{\arraystretch}{1.75}
\begin{tabular}{|m{2.5cm}|m{4.5cm}|m{8.5cm}|}
\hline
& \textbf{SQL Database} & \textbf{NoSQL Database} \\
\hline
\hline
Schema & Relations (structured or fixed types) & Non-structured and varied types\\
\hline
Storage Models & tables (in rows or records) & Combination of varied types\\
\hline
Scalability & lowly scalable & highly scalable\\
\hline 
ACID & Guaranteed & Supported by some of them \\
\hline
Programming Interfaces & Common (e.g., using SQL) & Varied APIs depending on databases \\
\hline
Examples & IBM DB2, Oracle, Microsoft SQL server, MySQL, Postgres, SQLite & Dynamo \cite{DeCandia:2007}, 
BigTable \cite{Chang:TCS2008}, Cassandra \cite{Lakshman:2009}, Hbase \cite{HBase:2016}, MongoDB \cite{Chodorow:2010}, Megastore \cite{Megastore:2011} and Spanner \cite{Corbett:2013} \\
\hline
\end{tabular}
\label{tab:sql-nosql}
\end{table*}

It is worth mentioning that the wireless devices are interconnected together to form the storage infrastructure for large scale wireless networks as shown in Fig. \ref{fig:storage}. There are several ways of managing the network infrastructure of storage systems: (i) Directed Attached Storage (DAS), (ii) Network Attached Storage (NAS) and (iii) Storage Area Network (SAN). Essentially, the distributed storage systems often introduce redundancy to increase reliability with erasure coding repair schemes \cite{Weatherspoon:2002}, which however inevitably cause the extra network traffic. There are a number of solutions proposed to address this challenge \cite{Zhu:IEEETComp15,CAChen:IEEETCom15,Chen:IEEETCom16,Jun:TCS2016,Sathiamoorthy:TMC2014,ALAWAMI:2017}. 

\subsection{Data management software}
\label{subsec:dataman}

Data management software plays an important role in constructing the scalable, effective, reliable storage system to support BDA. As shown in Fig. \ref{fig:storage}, we categorize the data management software into three layers: (i) file systems, (ii) database management systems and (iii) distributed computing models, which are explained as follows.

\emph{(i) File systems}

We start the introduction from simple file systems at a single personal computer (PC) to distributed file systems. In a PC, a file system is mainly used for the purpose of data storage and retrieval. The typical file systems for PCs include: 
a) The family of File Allocation Table (FAT) file systems: FAT 12, FAT 16, FAT 32; b) NTFS (New Technology File System) offering better security than FAT; c) index-node file system: Unix File Systems (GPFS, ZFS, APFS, etc.) and Linux File Systems (ext2, ext3, ext4); d) Contiguous allocation file systems: ISO 9660:1988 and Joliet ("CDFS"), which are mainly used for CD-ROM and DVD-ROM disks.

In order to support the file sharing and the collaboration, there are a number of distributed file systems including Andrew File System (AFS) and Network File System (NFS) \cite{ArpaciDusseau14-Book}, etc. However, most of these distributed file systems can only be accessed within a local area network and are not scalable to support the large scale data storage in wide area networks.

Google File System (GFS) \cite{Ghemawat:GFS2003} was proposed by Google to support the large data intensive applications (e.g., search engine) in distributed environments. In particular, GFS divides the data into equal-size blocks (typically with 64MB per block), which have been stored on different machines with several copies to ensure the fault tolerance. However, GFS is suffering from a number of drawbacks, such as inefficiency with smaller files and the degraded performance with the increased number of random writes. It is shown in \cite{McKusick:2009} that Google has partially solved the above defects of GFS.

Hadoop Distributed File System (HDFS) \cite{Shvachko:2010} proposed by Apache uses GFS for reference. HDFS is designed to store massive data sets reliably and support big data applications. The main idea of Hadoop is to partition data and computation across many servers and to execute computations in the parallel manner. Essentially, HDFS offers the scalable storage infrastructure to support Hadoop computational tasks.
 
In addition to GFS and HDFS, there are other distributed file systems, such as XtreemFS \cite{Hupfeld:2008}, C\# Open Source Managed Operating System (Cosmos) proposed by Microsoft \cite{Chaiken:2008} and Haystack proposed by Facebook \cite{Beaver:2010}. Most of them can partially or fully support the storage of large scale data sets.

\emph{(ii) Database management systems}

Database management systems (DBMS) concern how to organize the data in an efficient and effective manner. We roughly categorize DBMS into two types: (i) traditional relational DBMS and (ii) non-relational DBMS. In short, we name traditional relational DBMS as Structured Query Language (SQL) Database since most of them can support SQL queries. Similarly, we name non-relational DBMS as No-SQL database. Table \ref{tab:sql-nosql} summarizes the main differences between SQL DBMS and NoSQL DBMS.

SQL databases have been a primary data management approach since 1970s. There are several typical examples of SQL databases including commercial databases, such as Oracle, Microsoft SQL server and IBM DB2, and other open-source alternatives (e.g., MySQL and PostgreSQL). Before operating on data, a \emph{schema} must be defined first. The schema usually defined tables, field types, primary keys, indexes, relationships, triggers and stored procedures. 

SQL databases usually store data in tables of records, resulting in the poor scalability. There is a need to partition the data and distribute the operation load among different nodes (i.e. the database servers) with the growth of data. However, it is challenging to allocate data in a distributed manner \cite{rahimi2010distributed}. One of benefits of SQL databases is that most of SQL databases can guarantee ACID (Atomicity, Consistency, Isolation, Durability) properties, which are crucial to many commercial applications. Besides, the standardized SQL also facilitates the application development process since most of programming languages can operate on databases through general programming interfaces (such as ODBC and JDBC). 

Different from SQL databases, there is no schema to specify the data design in NoSQL databases. Besides, NoSQL databases also support various types of data, such as records, text, and binary objects. Compared with traditional relational databases, most of NoSQL databases are usually highly scalable and can support the tremendous amount of data. Thus, NoSQL databases have received extensive attention recently especially in BDA. We next briefly review several typical NoSQL databases.

\begin{figure*}[t]
\centering
\includegraphics[width=15cm]{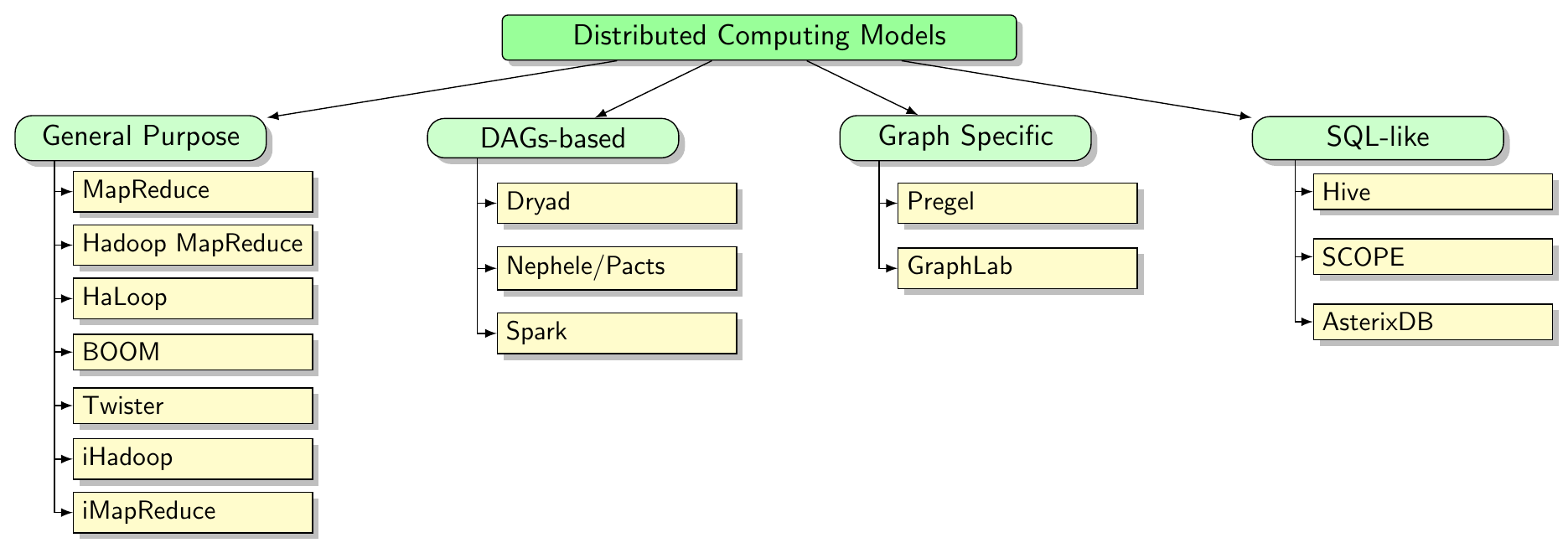}
\caption{Classification of distributed computing models}
\label{fig:class-dist}
\end{figure*}

\begin{itemize}
\item \emph{Key-Value databases}. In Key-Value databases, data is organized as key-value pairs. Similar to the primary key in SQL databases, each key in a Key-Value database is also unique. Through partitioning and replication mechanisms, such databases have higher scalability than traditional databases. Dynamo \cite{DeCandia:2007} is one of typical Key-Value databases though there are many other variants similar to Dynamo.

\item \emph{Column-Oriented databases}. Instead of storing data in rows, Column-Oriented databases store and manage data in columns. Bigtable \cite{Chang:TCS2008} invented by Google is one of the typical column-oriented databases. In Bigtable, data is organized in a sparse and distributed \emph{map}, in which column keys, row keys and time keys are the indexes. Bigtable is essentially implemented on GFS with the integration of other technologies. There are other alternatives to Bigtable, including Cassandra \cite{Lakshman:2009} and Hbase \cite{HBase:2016}. 

\item \emph{Document databases}. Document databases can support more complicated data types than SQL databases and key-value databases. Besides, there is no schema in document databases, consequently increasing the variety of data types. Typical document databases include MongoDB \cite{Chodorow:2010}, CouchDB \cite{Anderson:2010} and SimpleDB \cite{Chaganti:2010}.

\item \emph{Hybrid databases}. Hybrid databases integrate the benefits of SQL databases and No-SQL databases and obtain the better performance than both SQL databases and No-SQL databases. Megastore \cite{Megastore:2011} and Spanner \cite{Corbett:2013} are two of typical examples of hybrid databases. They achieve the high scalability of NoSQL databases while guaranteeing data consistency (e.g., ACID properties) of SQL databases.
\end{itemize}

\emph{(iii) Distributed computing models}

Though distributed databases \cite{rahimi2010distributed} were proposed to improve the performance of DBMS through distributing operations over distributed data storages, they cannot fulfill the growing demands of BDA on the tremendous amount of data in large scale distributed systems. As a result, there are a number of distributed computing models proposed for BDA. In this paper, we roughly categorize those models into the following three types as shown in Fig. \ref{fig:class-dist}.

\begin{itemize}
\item \emph{General Purpose Models} have been proposed to support BDA in large scale distributed database systems \cite{Koch2016}. Typical general purpose models include MapReduce \cite{Dean:2008}, a distributed programming model, is mainly used to big data analytics. MapReduce consists of \emph{Map} functions and \emph{Reduce} functions. Specifically, the \emph{Map} function first processes and sorts key-value pairs, and then save the intermediate data to temporary storage. The \emph{Map} function next consolidates the intermediate data (i.e., key-value pairs). Hadoop MapReduce \cite{Hadoop:MapReduce} is the open source implementation of Google MapReduce. One of limitations of MapReduce lies in the lack of iterations or recursions, which are however required by many data analysis applications, such as data mining, graph analysis and social network analysis. There are some extensions to MapReduce to address this concern, including HaLoop \cite{Bu:VLDB2010}, Berkeley Orders of Magnitude (BOOM) Analysis \cite{Alvaro:2010}, Twister \cite{Ekanayake:2010}, iHadoop \cite{Elnikety:CloudCom11} and iMapReduce \cite{Zhang:2012}. MapReduce has the merits including the scalability (e.g., it can be easily extended to a scenario of massive data sets and deployed in commodity PCs) and the simplicity (e.g., it can easily implemented by programmers without any parallel/distribute computing experience).

\item \emph{Directed Acyclic Graph (DAG)-based Models}. Dryad \cite{Isard:2007} models an application as a Directed Acyclic Graph (DAG), in which a vertex represents a computation and a directed edge represents a communication between vertices. In contrast to MapReduce, Dryad improves the flexibility while sacrificing model simplicity. Nephele/PACTs system \cite{Battre:socc2010} is a parallel data processing system including two components: Nephele and Parallelization Contracts (PACTs) where Nephele is a parallel computing system and PACT is a programming model. Nephele/PACTs has some similarity to Dryad owe to the flexible provision of dataflows on top of DAGs. Spark \cite{Zaharia:2010} is a cluster computing framework and can be easily extended to support different applications. Spark has the similar scheduling scheme to Dryad (based on DAGs), but it assigns tasks to machines based on the data locality information.

\item \emph{Graph Models}. Pregel \cite{Malewicz:2010} adopts a vertex-centric model, in which each vertex undertakes the computing tasks. Moreover, each edge consists of a source vertex and a destination vertex. In Pregel, every edge or every vertex is linked with a value. Similar to Pregel, GraphLab \cite{Low:2012} is also a vertex-centric approach. The main differences between Pregel and GraphLab are (i) different access privileges at vertices and edges and (ii) asynchronous/synchronous iterations. 

\item \emph{SQL-like Models}. SQL is a declarative language widely used in relational DBMS. Recently, there are a number of studies on designing distributed big data processing systems to support SQL or SQL-like language. In particular, Hive \cite{Thusoo:icde2010} developed by Facebook can undertake extensive data processing tasks. Hive supports SQL-like queries by using HiveQL. Essentially, Hive compiles HiveQL queries, packs them into MapReduce jobs and executes them on Hadoop. Hive has an excellent scalability and can be used to construct data storage systems while it has poor performance in processing interactive queries. Other alternatives include Structured Computations Optimized for Parallel Execution (SCOPE) \cite{Chaiken:2008} and AsterixDB \cite{Alsubaiee:2014}.
\end{itemize}

\begin{figure*}[t]
\centering
\includegraphics[width=16.5cm]{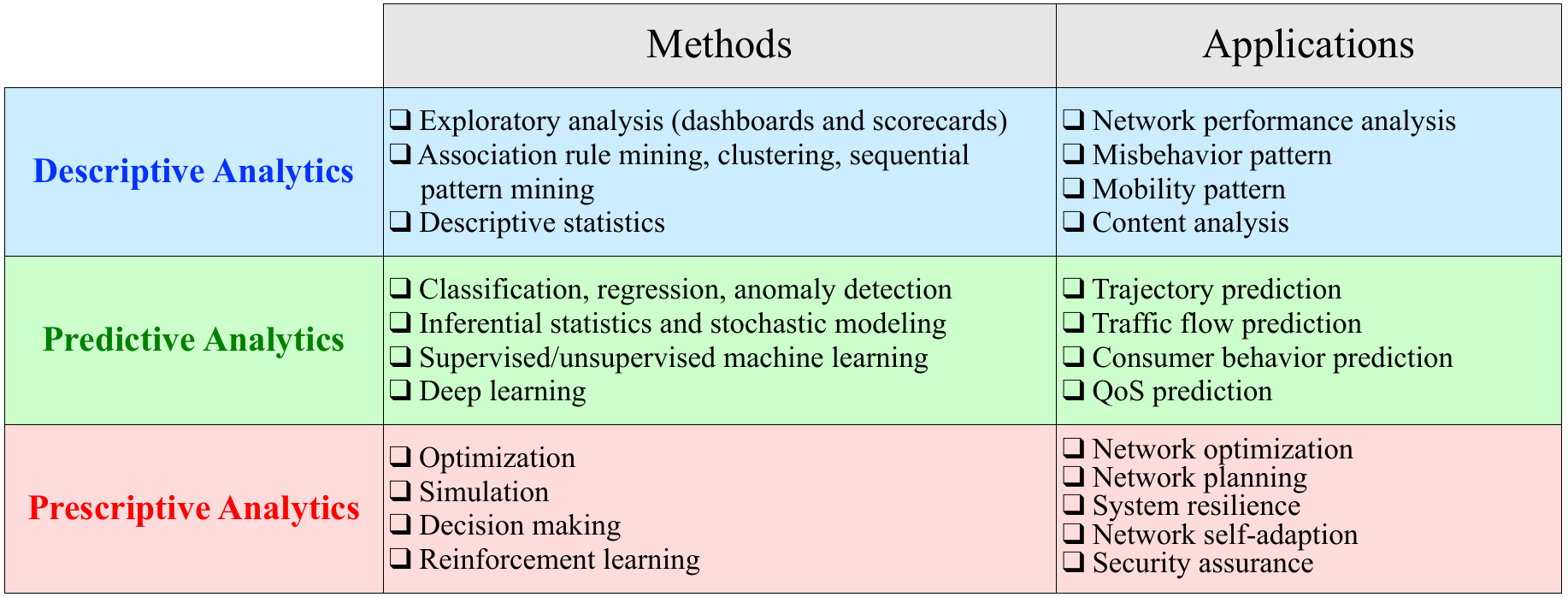}
\caption{Classification of Data Analytics Methods}
\label{fig:class-data-analysis}
\end{figure*}

\subsection{Opportunities in data storage}
\label{subsec:opp-storage}



Most of the challenges in data storage have been solved based on solutions in Section \ref{subsec:infrastructure} and Section \ref{subsec:dataman}. However, it is worth mentioning that there are still many issues not well addressed. We just identify three research opportunities as follows:
\begin{itemize}
\item \emph{Data storage and distributing computing models for heterogeneous networks.} There are few studies contributed to data storage and distributing computing models dedicated to different types of wireless networks. Different types of wireless networks may require different types of data storage and distributing computing models. 

\item \emph{Lightweight computing models.} Most of previous studies just assume that data is either stored at a cloud (or a fog) server or at a mobile device and sophisticated cloud-based computing models are used. However, many wireless devices (like IoT or sensors) may have the constraints such as limited storage and computational power. Heavy-weighted storage or computing models may not be feasible in this scenario.

\item \emph{Privacy preservation and security assurance.} Despite recent advances in distributed storage and computing systems, privacy-leakage concerns have received extensive attentions. For example, it is shown in \cite{Bazai:2017} that the output of MapReduce operations may cause the potential privacy leakage. To address the privacy leakage, Rao et al. \cite{RamMohanRao:2018} proposed a number of privacy-preservation techniques including data anonymity, data diversity, randomization and cryptographic schemes. Meanwhile, Wang et al. \cite{Wang:ICCCN18} proposed an efficient and verifiable erasure coding based storage to secure HDFS-like systems. In addition to these schemes, recent advances in blockchain technologies \cite{zibin2016blockchain} also show the strength in data privacy preservation and security assurance. 
\end{itemize}

\section{Data Analytics}
\label{sec:analysis}
The goal of data analysis is to extract useful information from large scale wireless networks. In this section, we first identify the challenges in data analytics in Section \ref{subsec:challenges_analytics}, then review the common solutions to these challenges in Section \ref{subsec:data_analysis}. In Section \ref{subsec:app_analysis}, we then enumerate the applications of the data analysis tools in the aforementioned typical wireless networks. We finally discuss the research opportunities in Section \ref{subsec:opp-analytics}. 

\subsection{Challenges in data analytics}
\label{subsec:challenges_analytics}

It is quite challenging in BDA for large scale wireless networks due to the tremendous volume, the heterogeneous structures, the high dimension and the wide data diversity.  The major challenges are summarized as follows.

\begin{itemize}
\item \emph{Data temporal and spatial correlation}. Different from conventional data warehouses, data of large scale wireless networks is usually spatially and temporally correlated. How to manage the data and extract valuable information becomes a new challenge. 

\item \emph{Efficient data mining schemes}. The tremendous volume of data of large wireless networks leads to the challenge in designing efficient data mining schemes due to the following reasons: (i) it is not feasible to apply conventional multi-pass data mining schemes due to the huge volume of data, (ii) it is critical to mitigate the data errors and uncertainty due to the erroneous features of wireless systems.

\item \emph{Privacy concern}. It is quite challenging to pertain the privacy of data during the analysis process. Though there are a number of conventional privacy-preserving data analytical schemes, they may not be applicable to the wireless data with the huge volume, heterogeneous structures, various types and spatio-temporal correlations. 

\item \emph{Real-time requirement}. Wireless data is often generated in real-time fashion. it is challenging to design and implement data analytics schemes in supporting real-time wireless applications while maintaining high performance (like prediction accuracy). 

\end{itemize}

We next survey the common data analytics methods to address these challenges.

\subsection{Common data analytics methods}
\label{subsec:data_analysis}

We categorize the common data analytics methods into the following types (as shown in Fig. \ref{fig:class-data-analysis}).

\subsubsection{Descriptive methods}

\emph{Descriptive analytics} mainly utilizes existing data sets to reveal the properties of data (i.e., what happened). We further categorize the descriptive methods into the following subcategories.

\begin{itemize}
\item \emph{Association Rule Mining} is to determine the dependence between two (or more than two) features. Typical association rule mining algorithms include Apriori algorithm and Frequent Pattern Growth (FP-Growth) algorithm \cite{Han:2012}. The main idea of Apriori algorithm is to identify the frequent individual items in the database and then to extend them to larger item sets as long as the frequency of the appearance of those item sets is sufficiently high in the database. One of disadvantages of Apriori algorithm lies the cost of generating candidate items. FP-Growth algorithm is based on an extended prefix-tree structure, which can store frequent patterns (hence this structure is also named as frequent-pattern tree).


\item \emph{Clustering} is a method of grouping objects such that objects in the same group have the higher similarity to each other than to those in other groups. The group consisting of similar objects is also named a \emph{cluster}. Typical clustering approaches include $k$-means and Density-based spatial clustering of applications with noise (DBSCAN) \cite{Han:2012}. The main idea of $k$-means is to divide $n$ objects into $k$ clusters such that objects with the nearest mean belong to the same cluster. DBSCAN partitions and associate the points into three groups: (1) the points, each of which contains a large number of neighbors; (2) the points that are reachable from the points from group (1); (3) the outliers that are not reachable from both the points in (1) and (2).

\item \emph{Sequential Pattern Mining} is the process that discovers relevant patterns between data examples where the values are delivered in a sequence \cite{Mooney:2013}. There are several typical sequential pattern mining algorithms: Generalized Sequential Pattern (GSP) and Sequential Pattern Discovery Using Equivalent Class (SPADE) and Prefix-Projected Sequential Pattern Mining (PrefixSpan) \cite{Han:2012}. GSP conducts multiple database passes. The first pass counts all single items (with sequence equal to 1). The second pass counts a set of candidate 2-sequences and one extra pass is conducted to identify their frequency. The set of candidate 2-sequences is used to generate the set of candidate 3-sequences. This process is repeated until no more frequent sequences are found. The main idea of SPADE is to divide the original problem into a number of sub-problems, which are independently solved in main-memory. During this process, efficient lattice search techniques are often used. PrefixSpan has further improved GSP and SPADE. In PrefixSpan, a sequence database is recursively partitioned into a number of sub-sets. Then, sequential patterns grow in each sub-database by selecting frequent segments only.

\item \emph{Descriptive statistics} can summarize features of data by exploiting measures of data. The typical measures of descriptive statistics include: (i) \emph{central tendency} such as mean, median and mode (e.g., bimodal and multimodal); (ii) \emph{dispersion} (or variability) including variance, covariance, tailedness (or kurtosis) and skewness \cite{Trochim:2016}. Besides, there are two typical descriptive statistical methods: \emph{univariate} analysis and \emph{bivariate} analysis. Univariate analysis mainly describes the distribution of a single variable while bivariate analysis is mainly used to the relationship between pairs of variables with sample data sets consisting of more than one variable.

\end{itemize}

\subsubsection{Predictive analytics}

\emph{Predictive analytics} mainly utilizes historical data to anticipate the trends of data (i.e., what will occur in the future). The predictive methods can be categorized into the following subcategories.

\begin{itemize}
\item \emph{Classification} is the process of assigning items in a collection to target categories. Classification is considered as an instance of supervised learning while clustering is considered as an example of unsupervised learning algorithms The typical classification algorithms include na\"{\i}ve Bayes, Bayes networks, $k$-Nearest Neighbors ($k$-NN), support vector machines (SVMs) and C4.5 \cite{Wu:2007}. Na\"{\i}ve Bayes is a simple scheme to classify features based on Bayes' theorem. Bayes classifier can minimize the probability of mis-classification. The main idea of $k$-NN is to classify an object by considering the closeness to its $k$ nearest neighbors. SVM is to find a hyperplane that maximizes the gap between the classes. C4.5 is an algorithm used to generate a decision tree, which can then be used for classification.

\item \emph{Regression} is a procedure of establishing a function to determine the relationship between target features. The regression is similar to classification though regression is mainly used to predict continuous values but classification is used to predict discrete values. Typical regression algorithms include linear regression and logistic regression. 

\item \emph{Anomaly Detection} (or outlier detection) is to identify objects that do not comply with an expected pattern as given. Anomaly detection approaches can be categorized into the following types \cite{Buczak:CST2016}: (a) \emph{supervised} anomaly detection schemes, (b) \emph{unsupervised} anomaly detection schemes, and (c) \emph{semi-supervised} anomaly detection schemes. The main difference between supervised schemes and unsupervised schemes mainly lies in the fact that the supervised schemes require a labeled (normal or abnormal) data set and a trained classifier. Semi-supervised schemes also use the trained classifier while it only consists of normal data.

\item \emph{Inferential statistics} can infer hidden distribution properties from the sample data. Different from descriptive statistics, inferential statistics deduces the data from a larger data set than the observed data set. A typical statistical inference procedure includes \emph{testing hypotheses} and \emph{deriving estimates} \cite{Bandyopadhyay:2011}. 

\item \emph{Stochastic modeling methods} have recently received extensive attention since they can capture the dynamic features of data traffic, predict user mobility and track objects. There are several typical stochastic models: dynamic Bayesian networks (DBNs), Markov models, Kalman filters and Extended Kalman filters \cite{Klaine:CST2017}. Most of these stochastic methods require collecting a certain amount of user data to provide stochastic models with parameter estimation (e.g., parameter estimation in transition matrices of Markov chains).

\item \emph{Supervised learning algorithms} require training data with labels first. Predefined
inputs and desired outputs are also given in supervised learning. The goal of supervised learning is to find the relationship between the inputs, the outputs and other arguments. Typical supervised learning algorithms include support vector machines (SVMs), na\"{\i}ve Bayes, Decision tree learning, $k$-Nearest Neighbors ($k$-NN), hidden Markov model, Bayesian networks, neural networks and Ensemble methods \cite{Russell:2009,Qiu:2016,Klaine:CST2017}. 

\item \emph{Unsupervised learning algorithms} do not require labeled training data and the desired outputs. The main goal of unsupervised learning is to classify items to target categories (i.e., clustering). Typical unsupervised learning algorithms include $k$-means, singular value decomposition (SVD) and Principal Component Analysis (PCA) \cite{abdi_2010}. The $k$-means algorithm is mainly used to assign data into different categories (or types). The main idea of PCA is to transform multiple correlated variables into new linearly orthogonal variables. These linearly uncorrelated variables is also called principal components.

\item \emph{Deep learning algorithms.} Conventional learning methods are mainly conducted in \emph{shallow-structured} learning architectures, which are not suitable for identifying complicated patterns. Recently, deep learning architectures have received extensive attention \cite{Sze:PIEEE2017}. There are two typical deep learning approaches: Convolutional Neural Networks (CNNs) and Deep Belief Networks (DBNs) \cite{Fadlullah:CST2017}.

\end{itemize}

\subsubsection{Prescriptive analytics}

\emph{Prescriptive analytics} extends the results of both descriptive and predictive analytics to make right decisions in order to achieve predicted outcomes (i.e., what should we do to achieve the goal?). The prescriptive methods can be categorized into the following subcategories.

\begin{itemize}
\item \emph{Simulation.} The proliferation of massive data of both descriptive and predictive analytics can be used to predict possible outcomes by mimicking possible scenarios with consideration of various constraints. Typical simulation tools include Monte Carlo simulation \cite{kroese2013handbook}, vehicular traffic flow simulations \cite{Treiber:2013Traffic}, Operator Training Simulator (OTS) for industrial systems \cite{Gerlach:processes2015}, etc.   

\item \emph{Optimization.} Optimization techniques include linear programming, integer programming, and nonlinear programming. The optimization problem usually concerns maximizing or minimizing a certain outcomes while satisfying given constraints. The typical optimal outcomes in wireless networks include throughput, delay, outage, energy consumption, operation cost and QoS \cite{kibria:2017bigdata}. 

\item \emph{Decision making.} A typical decision making process includes 1) identifying objectives (predictive outcomes),  2) giving a number of alternatives, 3) selecting the best alternative fulfilling the optimal outcomes via problem modeling (or simulation). Multiple criteria decision making (MCDM) has been typically used in decision making process. MCDM can help to find optimal outcomes in complicated scenarios with consideration of various criteria and conflicting objectives. Recently, MCDM models have been used in IoT systems \cite{Nunes:2016}, smart grids \cite{KUMAR:2017} and traffic flow control \cite{Jiang:ITS2018}.

\item\emph{Reinforcement learning algorithms} enable software agents to learn via the interactions with the context (or ambience) so that the cumulative reward can be maximized. One of the most popular reinforcement learning algorithms is Q-learning \cite{Russell:2009,Klaine:CST2017}, which is a model-free reinforcement learning technique. The main goal of Q-learning is to find an optimal policy on selecting actions for any finite Markov decision process. The procedure can be modeled by a quality value (i.e., Q-value). Q-learning has the strength that it works without the environment model. 

\end{itemize}

\subsection{Applications of data analytics}
\label{subsec:app_analysis}

We next offer an overview on the applications of the data analytics in large scale wireless networks.

\subsubsection{Mobile communication networks} 

BDA can be used to extract important features from mobile communication networks. We enumerate some of typical BDA applications in mobile communication networks as follows.

\begin{itemize}
\item \emph{Network performance analysis.} Classification and regression analysis can be exploited to analyze and predict the mobile traffic. For example, a machine learning based approach was proposed in \cite{ElKhayat:2010} to enhance the throughput of wireless networks via distinguishing packet loss causes. Moreover, the study of \cite{Zhang:TON15} proposed a new scheme named Robust statistical Traffic Classification (RTC) by combining both supervised and unsupervised ML techniques to address the {\it zero-day} problem. Furthermore, a multiclass-classification learning approach was proposed in \cite{Joung:CL2016} to solve the antenna-selection problem in mobile communications. 

\item \emph{Network security.} Both unsupervised clustering and supervised classification algorithms have been used to detect network intrusions or other malicious attacks. In particular, AdaBoost, Naive Bayes, Random Forest were proposed to detection the intrusion in 802.11 networks \cite{Kolias:CST2016}. Moreover, machine learning-based techniques can also be used in constructing intrusion detection systems for mobile clouds in heterogeneous networks \cite{gai2016intrusion}.

\item \emph{Mobility (Trajectory) prediction}. Mobility prediction has received extensive attention recently since it can offer the supports for various wireless services and mobile applications. There are a number of efforts been done in this area. Conventional approaches include using Markov models, Kalman filters and Extended Kalman filters to predict the user mobility. However, these approaches suffer from the poor accuracy. In \cite{xia2017:uptp}, a novel nonlinear SVM-based framework using massive spatio-temporal data was proposed. This approach was demonstrated to have the better accuracy than other existing methods. 
\end{itemize}

\subsubsection{Vehicular networks}

We enumerate several typical BDA applications in vehicular networks as follows.

\begin{itemize}
\item \emph{Misbehavior detection.} In particular, an intrusion detection systems based on deep neural networks was proposed to identify the malicious behaviors in vehicular networks \cite{kang2016intrusion}. Experimental results demonstrated the accuracy of the proposed approach. Moreover, in \cite{Scalabrin:2017}, a novel method based on Bayesian networks was proposed to analyze the anomaly in traffic data. 

\item \emph{Network performance.} The network topology and communication links in vehicular networks have experienced the frequent transition due to the high mobility of vehicles. The optimization of data transmissions in vehicular networks requires the accurate prediction of the moves of vehicles \cite{HYe:VTMag2018}. The study of \cite{Lai:DSN2015} proposed a novel routing information system called the machine learning-assisted route selection (MARS) system to estimate necessary routing information. Experimental results show that MARS can significantly improve the network performance.

\item \emph{Traffic prediction.} Traffic prediction in vehicular networks has received extensive attention recently. In \cite{YU:JTE2016}, a short-term traffic condition prediction model based on $k$-nearest neighbor algorithm was proposed. The study in \cite{Ko:Sensors16} proposed a Markov process based method to predict traffic conditions between roads. Moreover, a deep learning-based approach \cite{POLSON:2017,} was proposed to predict short-term traffic flow. Furthermore, Zhao et al. \cite{zhao2017lstm} proposed a novel method based on long short-term memory (LSTM) to predict the traffic flow. The benefit of this approach is the capability of capturing time-series features of traffic flow.

\end{itemize}

\subsubsection{Mobile Social Networks}

We enumerate several typical BDA applications of mobile social networks as follows.

\begin{itemize}

\item \emph{Community prediction.} Community activity prediction is useful to many applications, e.g., recommendation systems and network performance enhancement \cite{Rossetti2017}. In \cite{Puranik:2017}, two methods were proposed to analyze the dynamics of community structures. The study of \cite{Hao:SysJ2017} proposed an efficient algorithm of $k$-clique community detection using formal concept analysis (FCA). Experimental results show that the proposed algorithm has higher accuracy and lower computational cost than traditional methods.

\item \emph{Content analysis.} Social media contains a wide variety of data types from text, pictures, audio to video. As a result, the integration of multiple data analysis approaches together can improve the existing methods. In \cite{PENG:2017146}, the graph theory was used to investigate the influence of social content via a social relationship graph. 

\item \emph{Human behavior study.} Understanding human behavior via mobile social networks is extremely valuable for service providers and governors. In \cite{FXu:IEEENet16}, a study linking cyberspace and the physical world with social ecology via massive mobile cellular data is presented. Analytical results show that there are strong correlations between the mobile traffic and the human mobility; these results turn out to be related to social ecology.


\end{itemize}

\subsubsection{Internet of Things} 

We can apply data analysis approaches to extract valuable information from Internet of Things. We enumerate several typical applications as follows.

\begin{table*}[t]
\renewcommand{\arraystretch}{2.5}
\centering
\caption{Summary of solutions to challenges in BDA for large scale wireless networks}
\label{tab:summary}
\footnotesize
\begin{tabular}{|m{3.5cm}|m{3.5cm}|m{2.5cm}|m{2.5cm}|m{2.5cm}|}
\hline
\textbf{Challenges}  & \textbf{Mobile Communication Networks} & \textbf{Vehicular Networks} & \textbf{Mobile Social Networks} & \textbf{Internet of Things} \\
\hline\hline
\shortstack[l]{\\\textit{Data Acquisition}:\\$\cdot$ Data representation\\$\cdot$ Data collection\\$\cdot$ Data transmission}
   & \cite{AImran:IEEENet14} \cite{YHe:IEEEAccess2016}   \cite{Biral:DCN15} \cite{Su:IEEEComMag15}  \cite{Wang:IEEEComMag14} \cite{Checko:IEEECST15} \cite{Fan:IEEENet16} \cite{Mohajer:BDSON2017} \cite{Kuang:IEEENet16} \cite{Mehmeti:TMC2017} &  \cite{Ilarri:CST2015}  \cite{Mansour:2015} \cite{Brik:2016} \cite{placzek2017efficient}  \cite{Sahoo:ITS2017}  &    \cite{Richthammer:ARES13} \cite{leskovec2016snap} \cite{HAJARIAN2017} &  \cite{Kennedy:BigData2017} \cite{fahmy2016wireless}  \cite{ZChen:MIS2017}    \cite{JXu:IOTJ2017}    \cite{MEKKI2018} \cite{Iqbal:PIMRC18}\\
\hline

\shortstack[l]{\\\textit{Data Preprocessing}:\\$\cdot$ Integration\\$\cdot$ Duplication reduction\\$\cdot$ Cleaning and compression}
  &  \cite{Gubanov:ICDE2017} \cite{Zhang:SIGIR2002}  \cite{Sanctis:IEEENet2016} \cite{Fan:TKDE16}  &       \cite{Lenzerini:2002}  \cite{Hass:IBM02} \cite{Han:2012} \cite{XYu:ICC10}  \cite{Feldman:IPSN2012}  \cite{Fogue:TMC14} & \cite{Lenzerini:2002}\cite{Hass:IBM02} \cite{Aggarwal:2011} \cite{Zheng:www2010} \cite{Mehrotra:middleware2014}  \cite{JORGENSEN:2018}  &   \cite{Ertek:TSMCS2017}  \cite{Baba:2017}  \cite{Bhandari:sensors17}   \cite{MA:ESA2018} \cite{Tasnim:CCNC17} \cite{CDeng:IJPR2018}  \cite{ZZheng:TII18} \cite{Aleman:ICNC18}\\
\hline

\shortstack[l]{\\\textit{Data storage}: \\$\cdot$ Reliability \& Persistency \\$\cdot$ Scalability \\$\cdot$ Efficiency } &  \cite{Weatherspoon:2002}  \cite{CAChen:IEEETCom15}  \cite{Su:IEEEComMag15} \cite{ALAWAMI:2017} &   \cite{Sathiamoorthy:TMC2014} \cite{Li:tos2015}   &    \cite{Cheng:hpdc2015} \cite{ZSu:IEEENet16} \cite{Hu:JSAC2017}  &      \cite{Guerra:FAST2011}  \cite{Chen:IEEETCom16}  \cite{Jun:TCS2016} \cite{Sharma:Access2017}\\
\hline

\shortstack[l]{\\\textit{Data Analytics}:\\$\cdot$ Temporal-spatial correlation\\$\cdot$ Efficiency\\$\cdot$ Privacy\\$\cdot$ Real-time} &    \cite{ElKhayat:2010} \cite{Zhang:TON15} \cite{Joung:CL2016} \cite{Kolias:CST2016} \cite{gai2016intrusion}\cite{xia2017:uptp}  &  \cite{kang2016intrusion} \cite{Scalabrin:2017} \cite{YU:JTE2016} \cite{Ko:Sensors16}  \cite{POLSON:2017} \cite{zhao2017lstm} \cite{HYe:VTMag2018}  &  \cite{FXu:IEEENet16} \cite{Rossetti2017} \cite{Puranik:2017} \cite{Hao:SysJ2017} \cite{PENG:2017146}  &  \cite{PYChen:IEEE15} \cite{YZuo:2016} \cite{KDing:TSMCS2017} \cite{KHuang:TSMCS2017} \cite{Phoemphon2018} 
\cite{SaeediEmadi2018} \cite{ZZheng:TII18} \\
\hline
\end{tabular}
\end{table*}

\begin{itemize}
\item \emph{Event detection.} Event detection is one of the most important functions of WSNs. Applying data mining or machine learning approaches to WSNs can help to design effective event detection mechanisms. In particular, anomaly detection plays an important role in ensuring the reliability of industrial systems. In \cite{PYChen:IEEE15}, simulation results show that the distributed general anomaly detection scheme is scalable to large scale WSNs and outperforms other existing methods in terms of detection accuracy and efficiency. Moreover, the study of \cite{SaeediEmadi2018} proposed a DBSCAN-based scheme and an SVM-based scheme to detect anomalies in WSNs. Furthermore, it is crucial to extract events in the context of RFID data management applications. The work of \cite{KDing:TSMCS2017} proposed an RFID-enabled graphical deduction model to track objects with attritutes like time-sensitive state and position. Moreover, a machine learning based approach was proposed in \cite{Dagli:2015} to detect complex events in RFID systems.

\item \emph{Localization.} Localization is one of the most important functions of WSNs. Through localization, sensor nodes can determine the location of each other. There are two types of localization algorithms in WSNs: range-based algorithms and range-free algorithms. Range-based algorithms have relatively accurate localization while they require additional hardware (e.g., GPS) to obtain the distance information. Range-free algorithms require no extral hardware support but suffer from the poor localization accuracy. In \cite{Phoemphon2018}, a novel range-free localization algorithm based on the integration of Fuzzy Logic (FL) and Extreme Learning Machines (ELMs) was proposed. Experimental results demonstrate the effectiveness of the proposed scheme. One of the most important issues in IoT is the localization of tags and smart objects. However, the localization for IoT is more challenging than that in WSNs due to the following reasons: (i) GPS devices do not work indoor where RFID tags are typically used; (ii) RFID tags and IoT objects have the very limited power supply; (iii) RFID tags and IoT objects have more limited computational capability than nodes in WSNs. To solve the challenge, there are many machine learning based approaches proposed in the literature. For example, \cite{Kung:IJCS15} proposed a neural-network based RFID localization method, which uses the training data derived from both reference-tag coordinates and the coordinates generated by localization algorithms. Experimental results show that the proposed protocol can accurately locate critical objects.

\item \emph{Network optimization/security.} Due to the energy constraint of WSNs, how to design the network protocols to minimize the energy consumption is one of the challenges in WSNs. There are a number of efforts using machine learning methods to optimize the network performance of WSNs. For example, reinforcement learning based geographic routing algorithms were proposed in \cite{Alsheikh:2014}. Machine learning algorithms can also be used to improve the network security of IoT and WSNs. For example, \cite{KHuang:TSMCS2017} developed an efficient intrusion detection approach via learning traffic patterns. Experimental results show that this scheme has high detection accuracy. Moreover, machine learning approaches can also be used to design secure network protocol in WSNs\cite{AHAD:JNCA2016}.

\item \emph{Consumer behaviour prediction.} Consumer behaviour prediction plays an important role in many business application. In \cite{YZuo:2016}, a Bayesian network based approach was proposed to predict the customer purchase behaviour; the analysis is based on massive RFID data collected through RFID tags attached at customers. In \cite{ZZheng:TII18}, a novel method based on deep convolutional neural networks (CNN) was proposed to identify electricity theft (i.e., a malicious consumer behaviour) in smart grids.

\end{itemize}

\subsection{Opportunities in data analytics}
\label{subsec:opp-analytics}
%
%


Although many challenging issues as mentioned in Section \ref{subsec:challenges_analytics} have been partially or fully addressed, there are still many issues not well addressed.  We just enumerate some of research opportunities as follows:
\begin{itemize}
\item \emph{Energy-efficiency and time-sensitiveness}. For example, most of previous studies focus on the accuracy of data analytics. Many efforts have been taken on improving the analysis accuracy or reducing the analysis error. Few studies consider the practical issues such as energy-efficiency and time-sensitiveness when data analytics schemes are deployed at wireless nodes.
\item \emph{Privacy preservation in data analytics.} Due to the limited computational capability of some wireless nodes, many data analytics tasks are conducted at remote clouds, which are however owned by third parties. Many studies often ignore the key step in removing privacy-sensitive attributes from the collected data and directly upload the data to remote clouds. 
\item \emph{Security assurance in data analytics.} Although machine learning algorithms have shown their strength in data analytics in massive data, they are also suffering from vulnerabilities to malicious attacks. For example, it is shown in \cite{Wang:SP18} that hyper parameters in machine learning models can be stolen so as to breach proprietary rights and divulge confidential information. Moreover, machine learning models are vulnerable to malicious attacks such as  trojaning attack \cite{Liu:NDSS18} and poisoning attack \cite{Jagielski:SP18}. Hence, security countermeasures to remedy these vulnerabilities are expected. The recent advances in blockchain \cite{hndai:blockchain-iot2019} may help to improve the security of big data analytics.

\end{itemize}


\section{Future research directions}
\label{sec:open}


Table \ref{tab:summary} summarizes the solutions to challenges in big data analytics for large scale wireless networks in the aspects of data acquisition, data preprocessing, data storage and data analytics. Although many research challenges have been solved, there are still many research issues to be solved. We next discuss the future directions in big data analytics for large scale wireless networks.

\begin{figure*}[t]
\centering 
\includegraphics[width=16.5cm]{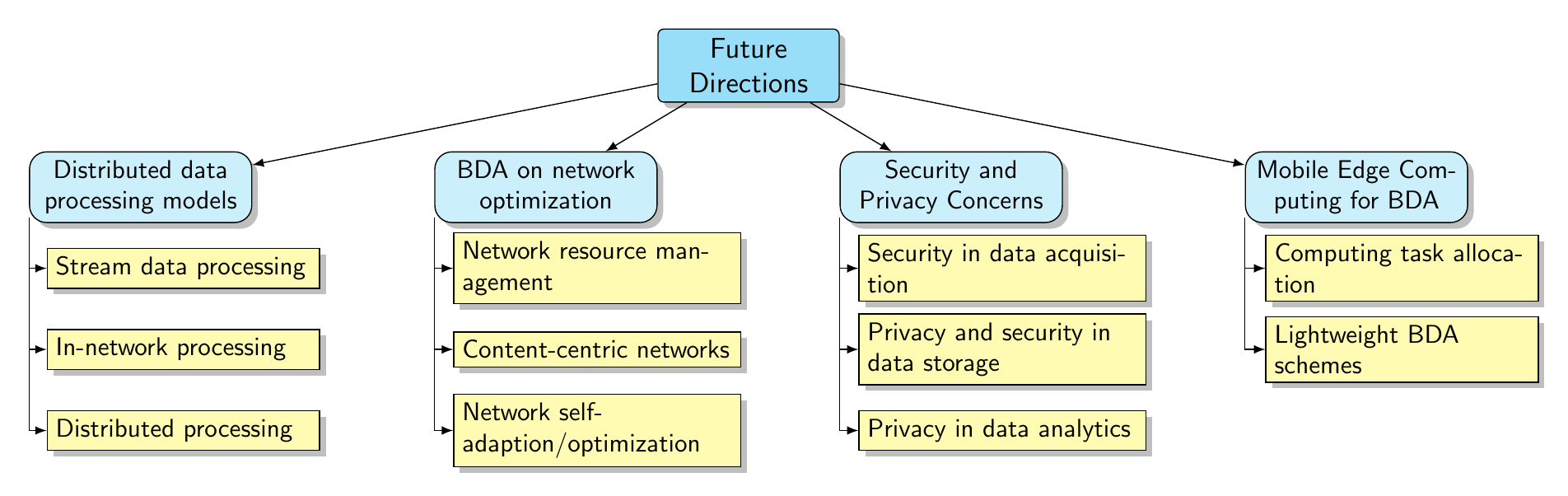}
\caption{Future research directions for BDA of wireless networks}
\label{fig:futuredir}
\end{figure*}

Fig. \ref{fig:futuredir} shows an overview on future directions in big data analytics for large scale wireless networks.

\subsection{Distributed data processing models}

Although a lot of efforts are done in developing distributed data processing models for large scale wireless networks, there are still many open research issues in this area. 

\begin{itemize}

\item \emph{Stream data processing}. Due to the tremendous volume of real-time data (e.g., sensor data from WSNs), it is impossible to store and process the entire data in memory. As a result, many conventional data analysis algorithms that require accessing the whole data sets do not work in this scenario. As mentioned in Section \ref{subsec:mcn-dp}, some data stream preprocessing schemes \cite{Aggarwal:2006} can be used in mobile communication networks and wireless sensor networks. However, there are few studies on data analysis of data streams as far as we know \cite{Krempl:kidd2014}.

\item \emph{In-network processing}. Since there are a large number of wireless nodes distributed in large scale wireless networks, the integration of the data generated from distributed nodes is necessary for data processing. However, the data fusion among the distributed networks inevitably causes significant communication cost. One of the solutions to this challenge is to conduct in-network processing among the whole network, in which data is processing at each node instead of at centralized servers. To further reduce the computational cost, it is usually advantageous to organize the nodes into clusters \cite{Diallo:TPDS15}. However, how to choose the cluster to fulfill the big data requirement becomes a new challenge.

\item \emph{Distributed processing}. Among the existing distributed data processing models, MapReduce \cite{Dean:2008} and its alternatives have many advantages, such as simple, fault tolerant and scalable. However, one of its limitations lies in the inefficiency compared with other parallel processing models. Conventional parallel computing models, such as Message Passing Interface (MPI), OpenMP (Open Multi-Processing), FPGA-oriented programming and GPU-based parallel processing (e.g., NVIDIA's CUDA), have the higher performance than MapReduce-like computing models. To integrate MapReduce-like models with parallel computing models can potentially improve the performance further. Besides, by exploiting the benefits of some dedicated processing platforms, we can further improve the existing data analysis algorithms. For example, there are some efforts in FPGA-oriented Convolutional neural network (CNN) \cite{Zhang:fpga2015} and FPGA deep learning algorithm \cite{LaceyTA:2016}, which have the better performance than those implemented in common PCs.

\end{itemize}

\subsection{Big data analytics on network optimization}

BDA can also help to optimize the network design of large scale wireless networks. We enumerate some of open issues on the impacts of BDA on network optimization as follows.

\begin{itemize}

\item \emph{Network resource management.} Based on BDA of network data, network administrators can predict the network resource demands. For an example as illustrated in \cite{KZheng:IEEENet16}, we can easily predict that there potentially exists a congested traffic in some locations of a city when a social event like a marathon takes place. Therefore, mobile operators can allocate more radio resources (i.e., more spectrum) to the hotspot so that the peak traffic can be absorbed smoothly \cite{KZheng:IEEENet16,CJiang:IEEEWC2017}.

\item \emph{Content-centric networks.} As suggested in many previous works \cite{SBi:IEEEComMag15,Su:IEEEComMag15,ZSu:IEEENet16}, to store some popular contents (also named caches) at base stations can significantly reduce the real-time traffic and consequently improve the network performance. However, how to determine the cache becomes a challenge. Essentially, we can acquire cache information through analyzing the application data. However, it is quite challenging to obtain the accurate user information due to the privacy preservation of user application data and the heterogeneous data types of various applications. 

\item \emph{Network self-adaption/self-optimization.} BDA is extremely useful in network self-adaption or self-optimization in self-organizing networks (SONs) \cite{Mohajer:BDSON2017,XWang:IEEEAccess15}. For example, Fan et al. proposed a self-optimization method with integration of a fuzzy neural network (NN) and reinforcement learning; this method can fulfill the requirements of coverage and capacity of SONs. In summary, more efforts shall be conducted in this new area.

\end{itemize}


\subsection{Security and Privacy Concerns}
\label{subsec:sec_privacy}

Security and privacy are important issues in BDA in wireless networks. Security and privacy are closely correlated while they are different from each other in the following aspects \cite{Abouelmehdi2018}: 1) Security is to ensure the \emph{confidentiality}, \emph{integrity} and \emph{availability} of data; 2) Privacy is to guarantee the proper usage of the data without the disclosure of user private information in the absence of user consent. We point out the future directions in security and privacy of BDA in wireless networks as follows.

\begin{itemize}
\item \emph{Security in data acquisition}. During this phase, the wiretapping behavior can take place anywhere and consequently leads to the information leakage. Therefore, substantial efforts on protecting the confidential information of wireless networks are necessary. Usually, we can apply encryption schemes in wireless networks \cite{Granjal:IEEECST2015}. However, it is infeasible to apply cryptography-based techniques in IoT due to the constraints of the energy and computational capability of smart objects \cite{Sadeghi:2015,Yang:iot17}. Therefore, new lightweight protection schemes are expected to be developed for IoT \cite{Liu:TDSC17}.

\item \emph{Privacy and security in data storage.} Once invasion on data storage system is successful, more personal confidential information can be disclosed. Thus, it is more critical to protect the stored data in this phase. Fortunately, it is easier to employ encryption algorithms to ensure security at data storage than that in data acquisition (transmission). However, it is still challenging to enforce the privacy-preserved operations in data storage \cite{BWang:TSC15}, especially when data storage service is offered by a third party \cite{JLin:IOTJ2017}. Mobile Edge Computing (MEC) \cite{Mach:CST17} can essentially offer a solution to the privacy-preservation in data storage by offloading the data from the untrusted third party to the trusted MEC server (deployed in proximity to the user).

\item \emph{Privacy in data analytics.} One of the major concerns of this phase lies in the balance between the privacy and the efficiency of data analysis \cite{RLu:IEEENet14}. For example, to protect the private user documents, usually the documents are encrypted and stored at a server (or a cloud). However, operations on the encrypted documents are time-consuming, which consequently lead to the in-efficiency in data analytics \cite{AU:2018}. There are still substantial efforts in the aspects of data publishing\cite{Zhang:TODS2017}, data mining output and distributed data privacy \cite{Mendes:Access17} needed to be done. 

\end{itemize}

\subsection{Mobile Edge Computing for BDA}
Due to inherent constraints such as limited power and inferior computational capability of wireless nodes, it is preferable to submit computing tasks from wireless nodes to remote cloud servers that have superior computational capability without resource constraints. However, cloud computing is also suffering from the limitations such as high latency, performance bottleneck, context unawareness and privacy exposure \cite{HLiu:IEEESys17}. MEC (or Fog computing) serves as a complement to cloud computing by overcoming the aforementioned limitations \cite{Tran:ComMag17}. The main idea of MEC is to offload computing tasks from remote clouds to various MEC servers deployed at base stations, IoT gateways and WiFi APs in a proximity to end users. In this way, the computing-intensive and delay-tolerant tasks will be executed at remote cloud servers while the delay-critical, computing less-intensive and context-aware tasks can be offloaded to edge servers. However, there are many challenges in MEC for BDA of wireless networks. 
\begin{itemize}
\item \emph{Computing task allocation.} There are various computing resources in wireless networks such as supercomputers at remote clouds, edge (fog) servers and mobile devices. It is necessary to determine how to allocate the computation resources at different computing devices. However, it is challenging to allocate and coordinate various computing resources distributed in large scale networks.
\item \emph{Lightweight BDA schemes.} It is shown \cite{YLin:ICLR2018} that AlexNet (i.e., a typical convolutional neural network) has the model size of 240MB, consequently resulting in huge communication cost from the server to the edge node. The resource limitation of edge servers and mobile devices motivates the research in designing lightweight BDA schemes and compressing BDA models \cite{CLeng:AAAI18}. It requires the efforts in hardware design, optimization, data compression, distributed computing and machine learning to achieve this goal \cite{YCheng:IEEESPM2018}. 
\end{itemize}

%
%
%
%
%
%
%
%
%
%
%
%
%

\section{Conclusion}
\label{sec:conc}

In this paper, we present a detailed survey on big data analytics (BDA) for large scale wireless networks. We first introduce the research methodology used in this paper. We then introduce data sources of several exemplary wireless networks including mobile communication networks, vehicular networks, mobile social networks, Internet of things. We next discuss the necessities and the challenges in BDA for large scale wireless networks. Based on our proposed four-stage life cycle of BDA for large scale wireless networks, we present a detailed survey on the existing solutions to the challenges in BDA. However, numerous research issues in this area are still open and need further efforts, such as improving distributed processing models, designing wireless networks with consideration of BDA and balancing the performance and privacy preservation trade-off in BDA. 

\section*{Acknowledgement}
The work described in this paper was partially supported by Macao Science and Technology Development Fund under Grant No. 0026/2018/A1. The authors would like to thank Gordon K.-T. Hon for his constructive comments.

\bibliography{IEEEabrv,BDAWirelessNets}

\end{document}